
\magnification=1200
\hsize=13cm


\def\ep{e^{2i\pi}}
\def\j#1{J^{#1}_i(z)}

\let \part\partial

\let \ga\gamma

\def\o#1#2{{#1\over#2}}

\def\bz{\bar z}

\def\ncomm#1#2{\left[~#1~,~#2~\right]}

\def\un{{\bf 1}}

\def\b{\beta}

\def\twovec#1#2{\left(\matrix{#1\cr #2\cr}\right)}

\def\pa{\partial}
\def \pab{{\bar \partial}}

\def\lam{\lambda}

\def\ep4{e^{i \pi /4}}

\def\emp4{e^{-i \o{\pi}{4}}}

\def\b{\beta}

\def\twovec#1#2{\left(\matrix{#1\cr #2\cr}\right)}

\def\pa{\partial}

\def\e2pi{e^{2\pi i}}
\def\em2pi{e^{- 2\pi i}}

\def\eip2n{e^{i\pi n/2}}




\def\ep{e^{2i\pi}}
\def\j#1{J^{#1}_i(z)}

\let \part\partial

\let \ga\gamma

\def\e2pi{e^{2\pi i}}
\def\em2pi{e^{- 2\pi i}}


\def\um{\o {1}{2}}

\def\o#1#2{{{#1}\over{#2}}}
\def\bz{{\bar z}}

\def\un{{\bf 1}}

\def\twovec#1#2{\left(\matrix{{#1}\cr {#2}\cr}\right)}

\def\ep4{e^{i \o{\pi}{4}}}
\def\emp4{e^{-i \o{\pi}{4}}}








\def\lam#1#2{\lambda^{{#1}}_{2}(z)}























\def\lam{\lambda}

\def\um{{\scriptstyle {\o{1}{2}}}}

\def\ep{e^{+}}
\def\em{e^{-}}

\def\pa{\partial}
\def\pab{\bar\pa}

\def\bei{\b_i}

\def\beit{\tilde{\b}_{i}}
\def\bit{\tilde{b}_i}

\def\bjt{\tilde{b}_j}

\def\j{{j}}






\def\vertenne#1#2#3{\exp \left [ \o{#1}{\sqrt{n^2 - 1}} \phi_0+
\o{#2}{\sqrt{2(n + 1)}}\phi_1 +
{\rm i}\o{#3}{\sqrt{2(n - 1)}} \phi_2 \right ] }


\def\parafer#1#2{\varphi^{#1}_{#2}}
\def\paraferbos#1#2{\exp \left [ \o{#1}{\sqrt{2(n + 1)}}\phi_1 +
{\rm i}\o{#2}{\sqrt{2(n - 1)}} \phi_2 \right ] }


\def\bosoncurrentn2#1{
\eqalignno{
T &=\o{1}{2} \left [ (\partial \phi_0 )^2 +
(\partial \phi_1 )^2 + (\partial \phi_2 )^2  \right ]
- \o{1}{2}Q_1^{(n)}\, \partial^2 \phi_1 \ \ , &(#1{\rm a})\cr
J & =
\sqrt{\o{n-1}{n+1}} \,\partial \phi_0 \ \ , &(#1{\rm b})\cr
G^{\pm}&=\sqrt{\o{2n-2}{n+1}} \, \Psi^{\pm} \, \exp \left [
\pm\sqrt{\o{n+1}{n-1}} \, \phi_0  \right] \ \ ,&(#1{\rm c}) \cr
\Psi^{\pm}&=\o{1}{\sqrt{2}} \, \exp \left [ \pm {\rm i} \sqrt{\o {2}
{n-1}} \, \phi_2 \right ] \, \left ( \sqrt{\o{n+1}{n-1}}
\partial \phi_1 \, \pm {\rm i} \,\partial \phi_2 \right )\ \ . &(#1{\rm d})\cr
}}


\def\bcbetaboson{{b \, =\, {\rm e}^{-\pi_1} ~~ , ~~ \beta \, =\,
{\rm e}^{{\rm i} \pi_2 - \pi_3 }~~ , ~~
c\, = \, {\rm e}^{\pi_1} ~~ , ~~ \gamma \, = \, {\rm e}^{-{\rm i}\pi_2 +\pi_3}
\partial \pi_3}}

\def\stresspi{
{T =\o{1}{2}\left [ (\partial \pi_1 )^2 +
(\partial \pi_2 )^2 + (\partial \pi_3 )^2 \right ]
- \o{1}{2} \left ( -\o{1}{n+1} \, \partial^2 \pi_1
+ {\rm i}\o{n}{n+1} \, \partial^2 \pi_2 - \partial^2 \pi_3
\right )\ \ . }}


\def\conversion#1{\eqalignno{
\pi_1 &=
\o{n}{\sqrt{n^2 - 1}} \, \phi_0 -
\o{1}{\sqrt{2(n+1)}}\,\phi_1
+ {\rm i} \o{1}{\sqrt{2(n-1)}} \, \phi_2 \ \ ,~ &(#1{\rm a})\cr
{\rm i} \, \,\pi_2 &=
\o{1}{\sqrt{n^2 - 1}} \, \phi_0
- \o{n}{\sqrt{2(n+1)}}\,\phi_1
+ {\rm i} \o{n}{\sqrt{2(n-1)}} \, \phi_2 \ \ ,~ &(#1{\rm b})\cr
\pi_3 &=
- \sqrt{\o{n+1}{2}}\,\phi_1
+ {\rm i} \sqrt{\o{n-1}{2}} \, \phi_2 \ \ .~ &(#1{\rm c})\cr
}}




\def\Xi{X^{i}}
\def\Xis{X^{i^{\star}}}

\def\Xjs{X^{j^{\star}}}
\def\posi{\psi^{i}}
\def\posis{\psi^{i^{\star}}}

\def\posjs{\psi^{j^{\star}}}
\def\posib{{\tilde \psi}^{i}}
\def\posisb{{\tilde \psi}^{i^{\star}}}
\def\posjb{{\tilde \psi}^{j}}
\def\posjsb{{\tilde \psi}^{j^{\star}}}
\centerline{\bf 1. Introduction}
\vskip 0.2cm

A widely studied class of heterotic string [1] vacua is provided by the
(2,2)-superconformal theories with central charge
$c=9$ [2--6] that correspond to space-time
compactifications on Calabi-Yau manifolds [7--11]. Of major interest
in this context is the special K\"ahler
geometry [12--16]
of the associated Calabi-Yau moduli spaces [17--20], that to a large
extent determines all the relevant features of the low-energy effective
supergravity action [21].

Recently it has become evident that there is a deep relation between
this special K\"ahler
geometry  and the {\it flat} geometry of the
topological field theories [22--29]
obtained by twisting and deforming N=2 superconformal models
[30--32].
So far
the analysis of these issues has been based mainly
on the use of the Landau-Ginzburg
formulation of topological models and of its relation with
singularity theory [23-26,33-36].
In such a formulation the parameters entering the
Landau-Ginzburg superpotential
are interpreted as coordinates of some (moduli) space;
however these
are not the {\it flat} coordinates one is interested in,
since they do not
correspond to deformations around the conformal point;
rather they are
related to the latter by the solution of a uniformization problem,
which in general involves higher transcendental
functions.

In this paper we present an alternative approach to topological models
where the relation to singularity theory is directly obtained
in a natural system of flat coordinates. At the same time, these are
the parameters of a Landau-Ginzburg superpotential as well as the
deformations around the conformal point.
The first step of our construction is the use of free first-order
$(b,c,\beta,\gamma)$-systems to
describe N=2 superconformal theories as proposed in [37]. We then
show that an arbitrary interaction of the Landau-Ginzburg type -- {\it i.e.}
characterized by a polynomial potential $V$ -- can
be added to the free Lagrangian without spoiling the superconformal
invariance if $V$ is a quasi-homogeneous function.
The deformation parameters of the potential
are then shown to be the flat coordinates.

The paper is organized as follows. In Section 2 we discuss the
classical $(b,c,\beta,\gamma)$ realization
of N=2 superconformal models and
their topological twisting, and compare our approach
with the conventional topological Landau-Ginzburg description.
In Section 3 we discuss the
quantum properties of our
theories and show that the
classical (2,2)-superconformal invariance
extends trivially to the quantum
level due to the absence of loop corrections.
In Section 4 we bosonize the first-order systems and make
contact with the usual Coulomb gas
representation of N=2 minimal models
[38,39].
Section 5 contains explicit calculations of topological correlation
functions and a direct verification that
our coordinates are indeed the
flat ones. We retrieve established results for the minimal
models with $c<3$ and for the torus with
$c=3$. In Appendix A we give a short derivation of the Landau-
Ginzburg action using rheonomic approach, and in Appendix B
we show how our first
order lagrangian can be seen as a BRST commutator, with a suitable
choice of BRST charge.
\vfill \eject

\centerline{\bf  2.  Lagrangian formulation of N=2 theories with
$(b,c,\beta,\gamma$)-systems}
\centerline {\bf and comparison with the Landau-Ginzburg approach}
\vskip 0.3cm

In this section we consider the realization of (2,2)-supersymmetric
models in terms of free $(b,c,\beta,\gamma)$-systems
recently introduced in [37], and generalize it to include
interactions of the Landau-Ginzburg (LG) type.
As stressed in the introduction, N=2 superconformal theories
with $c=9$ are of primary interest in connection with
superstring compactifications on  six-dimensional Calabi-Yau spaces.
More generally, a (2,2)-superconformal theory with $c=3d$
corresponds to the critical point of an N=2 $\sigma$-model
on a target space with
complex dimension $d$ and vanishing first Chern class. We will call
these spaces Calabi-Yau $d$-folds.

In the LG formulation of (2,2)-supersymmetric
models, the superconformal theory
is viewed as the infrared fixed point of a two-dimensional N=2 Wess-Zumino
model
with a polynomial superpotential $W$.
In particular, when $W$ is an analytic quasi-homogeneous
function of the chiral superfields $X_i$, one can recover the complete
ADE classification of the N=2 minimal models
from the ADE classification of quasi-homogeneous potentials
with zero modality [5,33-36].
Furthermore, the polynomials $W$'s can be identified with
those used in the construction of
Calabi-Yau $d$-folds. Indeed,
it can be shown [5,35] that a superconformal model with $c=3d$,
corresponding to a LG potential $W$, is the same as that
associated to a $\sigma$-model on the Calabi-Yau
$d$-fold defined by the polynomial constraint
$W(X_i)=0$ in a suitable projective
or weighted projective space \footnote{$^1$}{In general
$W(X_i)$ is actually the sum of several terms
$W(X_i)=\sum\limits_\alpha
W_{\alpha}(X_i)$ and the Calabi-Yau $d$-fold is given by the
complete intersection $W_1=W_2=\cdots=W_n=0$ [5].}.

In this section we
show that it is possible to add a polynomial
interaction $V$ of the LG type to
a collection of free first-order
$(b,c,\beta,\gamma)$-systems in such a way that, if $V$ is a
quasi-homogeneous function,
the theory possesses an N=2 superconformal symmetry
already at the classical level.
We also show that the interaction potential
unambiguously fixes the weights of the pseudo-ghost fields.
As in the standard LG case, also here we can
recover the ADE
classification of the N=2 minimal models from ADE
classification of the interaction potential [35,36]; however in our case
the theory is always manifestly superconformal
invariant.
Our formulation allows us to
add all relevant perturbations (versal deformations of the
potential) and to study the renormalization group flows in a
very simple way.
Whenever we use a quasi-homogeneous potential
with modality different from zero, we can
study marginal deformations and eventually
Zamolodchikov's metric on the associated
moduli space. Alternatively, we can consider topological
models by ``twisting'' the generators of the
superconfromal algebra and compute topological correlation
functions. Our formulation provides valuable
methods to evaluate these latter.

We start this program by defining our model.
We consider a collection of pseudo-ghost fields
$\{b_\ell,c_\ell,\beta_\ell,\gamma_\ell;{\tilde{b}}_r,
{\tilde{c}}_r,{\tilde{\beta}}_r,{\tilde{\gamma}}_r\}$ where
$\ell=1,\dots,N_{\rm L}$ and $r=1,\dots,N_{\rm R}$. $\beta_\ell$
and $\gamma_\ell$ form a bosonic first-order
system with weights $\lambda_\ell$ and $1-\lambda_\ell$ respectively
whereas
$b_\ell$ and $c_\ell$ form a fermionic first-order system
with weights $\lambda_\ell+\um$ and $\um-\lambda_\ell$ respectively.
The same can be said for the tilded fields with $\lambda_\ell$
replaced by ${\tilde{\lambda}}_r$. The action
is
$$
S = \int d^2 z ~ {\cal L} = \int d^2z ~
\left( {\cal L}_0 + \Delta {\cal L}\right )
\eqno(2.1)
$$
where
$$\eqalign{
{\cal L}_0 =~~&\sum_{\ell}\left[
-\lam_\ell\, \beta_\ell \bar \partial \gamma_\ell +
(1- \lam_\ell)\,\gamma_\ell \bar \partial \beta_\ell
- \left(\lam_\ell +\um\right)\, b_\ell \bar \partial c_\ell
+\left(\lam_\ell -\um\right)\,
c_\ell \bar \partial b_\ell\right]\cr
+&\sum_{r}\left[-
{\tilde \lam}_r \,\tilde\beta_r  \partial \tilde \gamma_r
+(1- {\tilde \lam}_r)\,
\tilde\gamma_r  \partial \tilde \beta_r
- ({\tilde \lam}_r +\um)
\,\tilde b_r \partial
\tilde c_r  + ({\tilde\lam}_r -\um) \,
\tilde c_r
\partial \tilde b_r\right]
}\eqno(2.2{\rm a})
$$
and
$$
\Delta {\cal L}  = \sum_{\ell,r}
b_\ell \, \tilde b_r\,
\partial_\ell V(\beta) \,\tilde \partial_r \tilde V(\tilde \beta)\ \ .
\eqno(2.2{\rm b})
$$
Here and in the following we use the short-hand notations
$\pa_\ell \equiv {\pa}/{\pa \beta_\ell}$ and
$\tilde \pa_r \equiv {\pa}/{\pa
\tilde \beta_r}$.
${\cal L}_0$ in (2.2a) represents the standard free
Lagrangian for first-order systems of
the given weights and $\Delta{\cal L}$
in (2.2b) defines
an interaction of the LG type when $V$ and $\tilde V$ are
polynomial functions of $\beta_\ell$ and $\tilde \beta_r$ respectively.

{}From (2.1) one can derive the following equations of motion
$$
\eqalign{
\bar\pa \beta_\ell & = 0 \ \  , \ \ \bar\pa b_\ell = 0\ \ , \cr
\pa \tilde \beta_\ell & = 0
\ \ , \ \ \pa \tilde b_\ell=0 \ \ ,\cr
\bar\pa c_\ell &= \sum_{r}
\tilde b_r \pa_\ell V(\beta) \tilde\pa_r
\tilde V(\tilde \beta)\ \ ,\cr
\bar\pa \gamma_\ell &= \sum_{m,r}
b_m \tilde b_r \pa_\ell
\pa_m V(\beta) \tilde\pa_r
\tilde V(\tilde \beta)\ \ ,\cr
\pa \tilde c_r &= -\sum_{\ell}b_\ell \pa_\ell V(\beta)
\tilde\pa_r \tilde V(\tilde \beta)\ \ ,\cr
\pa \tilde \gamma_r &= \sum_{\ell,s}
b_\ell \pa_\ell
V(\beta)\tilde b_s \tilde\pa_r
\tilde\pa_s \tilde V(\tilde \beta) \ \ .
\cr}\eqno(2.3{\rm a})
$$
The first two lines of (2.3a) show that $\beta_\ell$,
$b_\ell$, $\tilde \beta_r$ and $b_r$ satisfy the same equations
as in the free case,
whereas $c_\ell$, $\gamma_\ell$,
$\tilde c_r$ and $\tilde \gamma_r$ have no longer a definite
holomorphic or anti-holomorphic character in the
presence of the interaction.
We can write formal solutions to the equations
for $c_\ell$ and $\gamma_\ell$ as follows [40]:
$$\eqalign{
c_\ell(z,\bar z) &=c_\ell^0(z) +\int_{\Delta} \o{d^2w}{2 \pi {\rm i}}
\o{1}{w-z}\sum_{r}
\tilde b_r(\bar w) \partial_\ell V(\beta
(w)) \tilde\pa_r\tilde V(\tilde \beta (\bar w))\ \ ,\cr
\gamma_\ell(z,\bar z) &=\gamma^0_\ell (z) +\int_{\Delta} \o{d^2w}{2 \pi {\rm
i}}
\o{1}{w-z}\sum_{m,r} b_m(w)
\tilde b_r(\bar w) \partial_\ell \pa_m V(\beta
(w)) \tilde\pa_r \tilde V(\tilde \beta (\bar w))\ \ ,\cr }\eqno(2.3{\rm b})
$$
where $\Delta$ is a disk containing $w$, and $c_\ell^0$and
$\gamma_\ell^0$
are arbitrary holomorphic fields.
Similar formal expressions (with the obvious changes)
hold also for
$\tilde c_r$ and $\tilde \gamma_r$.
It is fairly easy to realize that
under canonical quantization of (2.1)
the fundamental operator product
expansions are the same as in the free case. Indeed,
even in the presence of the interaction, we have
$$
\eqalign{
\beta_\ell(z) \,\gamma_m(w,\bar w)&=- \o{\delta_{\ell m}}{z -w}
+ \ \cdots \ \ , \cr
b_\ell(z)\,c_m(w, \bar w) & =\o{\delta_{\ell m}} {z-w} + \ \cdots \ \ ,
\cr}\eqno(2.4)
$$
and similarly for the tilded fields.
Of course, the interaction is not immaterial and it
has to be carefully analyzed in a complete quantum treatement as we
will do in Section 3.

It is well-known that ${\cal L}_0$ in (2.2a) describes a
(2,2)-superconformal field theory with central charges
$$
c_{\rm L} = \sum_{\ell} (3 - 12 \lambda_\ell) ~~~~,~~~~
c_{\rm R} = \sum_{r} (3 - 12\tilde \lambda_r) \ \ ,
\eqno(2.5)
$$
for the left and the right sectors respectively.
We will now show that the addition of the interaction $\Delta {\cal L}$
does not destroy this (2,2)-superconformal invariance
if $V$ and $\tilde V$ are quasi-homogeneous functions, {\it i.e.}
if for any $a \in {\bf R}^+$,
$$
V(a^{\omega_\ell} \beta_\ell) = a \, V(\beta_\ell) ~~~~,~~~~
\tilde V(a^{\tilde \omega_r} \tilde\beta_r) = a \,
\tilde V(\tilde\beta_r) \ \ .
\eqno(2.6)
$$
The parameters $\omega_\ell$ and
$\tilde \omega_r$ are called the homogeneous weights
of $\beta_\ell$ and $\tilde \beta_r$ respectively.
By enforcing the requirement that the interaction Lagrangian
$\Delta {\cal L}$ have the correct dimensions, one can
see that
$$
{\omega_\ell} = 2 \lam_\ell ~~~~,~~~~
{\tilde \omega_r} = 2 \tilde \lambda_r \ \ ;
\eqno(2.7)
$$
the parameters $\lambda_\ell$ and $\tilde \lambda_r$ of
the free Lagrangian (2.2a) are therefore {\it fixed}
by the interaction terms.
When (2.6) and (2.7) are satisfied, the action $S$ in (2.1)
is invariant under the
following N=2 holomorphic supersymmetry transformations
$$
\eqalign{
\delta \beta_\ell &=2\sqrt{2} \epsilon^- b_\ell \ \ ,\cr
\delta b_\ell &={1\over {\sqrt{2}}} \epsilon^+ \pa\beta_\ell
+\sqrt{2}\lam_\ell \,\pa \epsilon^+
\beta_\ell \ \ ,\cr
\delta c_\ell &=2\sqrt{2}\epsilon^- \gamma_\ell \ \ ,\cr
\delta \gamma_\ell &={1 \over {\sqrt{2}}}
\epsilon^+ \pa c_\ell -\sqrt{2}\left(\lam_\ell-\um
\right) \pa \epsilon^+c_\ell \ \ ,\cr
\delta \tilde \beta_r &=0\ \ ,\cr
\delta \tilde b_r &=0 \ \ ,\cr
\delta \tilde c_r &=-\o{1}{\sqrt{2}} \epsilon^+ V(\beta) \tilde\pa_r
\tilde V(\tilde
\beta) \ \ ,\cr
\delta \tilde \gamma_r &=\o{1}{\sqrt{2}} \epsilon^+ V(\beta)
\sum_{s}\tilde
\pa_r \tilde\pa_s \tilde V(\tilde\beta) \tilde b_s \ \ ,\cr}
\eqno(2.8)
$$
where $\epsilon^\pm$ are arbitrary holomorphic functions
($\bar\pa
\epsilon^{\pm}=0$).
The action $S$ is also invariant under
N=2 anti-holomorphic symmetries which are
similar to the ones defined in (2.8), with the exchange of the
tilded and untilded quantities, and the replacement
of $\epsilon^\pm$ with arbitrary
anti-holomorphic functions $\bar \epsilon^\pm$
($\pa\bar\epsilon^{\pm}=0$).
Moreover, if we relax the hypothesis
that $V$ and ${\tilde V}$ are quasi-homogeneous,
the transformations (2.8) and their $\bar \epsilon$-analogues remain
symmetries of (2.1) provided
$\epsilon^{\pm}$ and ${\tilde \epsilon}^{\pm}$ are constant
parameters.
This means that our model has a
global N=2 supersymmetry for any choice of
$V$ and $\tilde V$,
and an N=2 superconformal invariance
for quasi-homogeneous potentials.

The conserved Noether's currents associated to the
symmetries (2.8) are \footnote{$^2$}{
{}From now on, to avoid repetitions we will discuss only the left sector
and understand that similar considerations can be made in the right sector,
with some obvious change of signs.}
$$\eqalign{
G^+_z &=\sqrt{2}
\sum_{\ell}
\left[\left(\um-\lam_\ell\right)
c_\ell \pa \beta_\ell -\lam_\ell \,\beta_\ell \pa c_\ell \right]
\ \ ,\cr
G^+_{\bar z} &= \sqrt{2}
\sum_{\ell}\left[
\lam_\ell \beta_\ell \bar\pa c_\ell
+\left(\lam_\ell - \um\right)\bar\pa
\beta_\ell c_\ell \right]-\o{1}{\sqrt{2}}
\sum_{r}
V(\beta)
\tilde b_r \tilde \pa_r \tilde V(\tilde \beta) \ \ ,\cr
G^-_{z} &=2 \sqrt{2}
\sum_{\ell}\gamma_\ell b_\ell \ \ ,\cr
G^-_{\bar z} &=0 \ \ .\cr}
\eqno(2.9)
$$
If we use the equations of motion (2.3a)
for quasi-homogeneous potentials, we see that
$G^+_{\bar z}$ vanishes on-shell; thus from the
conservation laws we deduce that
$G_z^+$ and $G_z^-$ are
holomorphic currents even if they contain
the non-holomorphic fields $c_\ell$ and $\gamma_\ell$.
We denote these currents by $G^\pm(z)$.

The action (2.1) is also invariant under holomorphic
conformal reparametrizations and $U(1)$-rescalings of the
fields; the conserved Noether's currents
associated to such symmetries are the stress-energy
tensor $T_{\mu\nu}$ and the $U(1)$-current $J_\mu$.
For homogeneous potentials it is not difficult to see that
the trace of $T_{\mu\nu}$ and the ${\bar z}$-component
of $J_\mu$ are zero on-shell (see also Section 3).
Therefore, from the conservation laws, we deduce that
$$
T_{zz} =
\sum_{\ell}
\left[-\lam_\ell\b_\ell\pa\ga_\ell+(1-\lam_\ell)\ga_\ell\pa\b_\ell
-\left(\lam_\ell+\um\right)b_\ell\pa c_\ell +
\left(\um-\lam_\ell\right)c_\ell\pa b_\ell\right] \ \ ,
\eqno(2.10)
$$
and
$$
J_z = \sum_{\ell}
\left[(2\lam_\ell-1) b_\ell c_\ell+2\lam_\ell\b_\ell\ga_\ell
\right]
\eqno(2.11)
$$
are holomorphic currents. We denote them by
$T(z)$ and $J(z)$ respectively.

Using the OPE's in (2.4), it is straightforward to check
that $T(z)$, $G^\pm(z)$ and $J(z)$ close an N=2 superconformal
algebra with central charge
$$
c_{\rm L} = \sum_{\ell} (3 - 12 \lambda_\ell) \ \ .
\eqno(2.12)
$$
Thus, we have shown that the interaction $\Delta {\cal L}$ with
homogeneous polynomials $V$ and $\tilde V$ does not
spoil the superconformal properties of ${\cal L}_0$.

In our formulation the ADE classification of N=2 superconformal models
is an immediate consequence of ADE classification
of homogeneous polynomials of zero modality [5,35,36]. The latter are
$$
\eqalign{
A_{n}&:\quad V =\o{1}{n+1}\beta^{n+1}~~~\Rightarrow \lambda=\o{1}{2n+2}
\quad n\ge 1
\ \ ,\cr
D_{n}&:\quad V =\o{1}{n-1}\beta_1^{n-1}+\o{1}{2}\beta_1 \beta_2^2 ~~~
\Rightarrow
\lambda_1 =\o{1}{2n-2} \,\, , \, \lambda_2=\o{n-2}{4n-4}\quad n\ge 2\, ,
\cr
E_6&:\quad V=\o{1}{3}\beta_1^3 +\o{1}{4}\beta_2^4  ~~~\Rightarrow
\lambda_1=\o{1}{6} \,\, , \, \lam_2=\o{1}{8}\ \ ,\cr
E_7&:\quad V=\o{1}{3}\beta_1^3 +\o{1}{3}\beta_1 \beta_2^3
{}~~~ \Rightarrow
\lambda_1 =\o{1}{6} \,\, , \,\, \lam_2=\o{1}{9} \ \ ,\cr
E_8&:\quad V =\o{1}{3}\beta_1^3+\o{1}{5}\beta_2^5
{}~~~ \Rightarrow
\lambda_1 =\o{1}{6} \,\, , \,\, \lam_2=\o{1}{10}\ \ . \cr}
\eqno(2.13)
$$
We remark
that the values of $\lambda_\ell$'s listed in (2.13) are
fixed by the homogeneous weights of $\beta_\ell$'s
according to (2.7). If we now insert such values into (2.12)
we obtain the correct central charges for the N=2 minimal models
in the ADE classification, namely
$$
c(A_{n})= \o{3n-3}{n+1}\,,c(D_{n})=\o{3n-6}{n-1}\,
,\, c(E_6)=\o{5}{2} \,,\, c(E_7)=\o{8}{3}\,,\,c(E_8)=\o{14}{5}\ \ .
\eqno(2.14)
$$
It is also interesting to observe that the ring determined by
the potential $V$, which contains all
polynomials in $\beta_\ell$'s modulo the vanishing
relations $\pa_\ell V=0$, coincides with the
ring of
chiral primary operators of the N=2 minimal model
associated to $V$. Indeed, using
(2.10) and (2.11), one can easily check that
 the $U(1)$-charge of $(\beta_\ell)^n$ is
 twice its conformal dimension.

In order to compare our formulation of N=2
supersymmetric models with the standard LG approach and to
establish a clear correspondence with the topological conformal
field theories, it is convenient to
specialize our system to the case
of a complete symmetry between the left and the right sectors
($N_{\rm L}=N_{\rm R}=N$).
We shall then
consider interactions of the form
$$
\Delta {\cal L}
= \sum_{i,j=1}^N b_i{\tilde b}_j \,{\partial_i}{\tilde
\partial_j} W
\eqno(2.15)
$$
where $W$ is a quasi-homogeneous function of the variables
$$
X_i \, = \, \beta_i{\tilde \beta}_i
{}~~~,~~~i=1,\dots,N \ \ .
\eqno(2.16)
$$
This is clearly a very special case of
(2.2b).
Under these conditions, the Lagrangian (2.1) describes the infrared fixed
point of an ordinary N=2 LG
model with superpotential $W$. This equivalence will
be fully illustrated in the next sections. Here instead,
we discuss the topological formulation of
our models. It is well known that given an N=2
superconformal algebra generated by $T(z)$, $G^\pm(z)$ and $J(z)$,
one obtains a topological conformal algebra by ``twisting''
the currents according to
$$
\eqalign{ {\hat T}_{\pm }(z)  & = T(z) \pm \o{1}{2}
\partial J(z) \ \ ,\cr
{\hat J }_{\pm}(z) & = \pm  J(z)  \ \ ,  \cr
Q_{\pm }(z) & =  G^{\pm}(z) \ \ , \cr
G_{\pm }(z)  & = G^{\mp}(z)\ \ .\cr }
\eqno(2.17)
$$
We refer the reader to the original literature [30-32] for a discussion of
the properties of the topological conformal algebra
generated by ${\hat T}_\pm$, ${\hat J}_\pm$, ${Q}_\pm$,
${G}_\pm$; here we
simply mention that $Q_\pm(z)$ is interpreted as a BRST current and
the cohomology classes of the BRST charge
$$ Q^{\rm BRST}_\pm = \oint dz ~Q_\pm(z)
\eqno(2.18)
$$
are identified with the physical fields of the topological
theory. The two choices of signs in (2.17) lead to two
different sets of BRST invariant states: the chiral primary
fields of the original N=2 superconformal algebra for
the $+$ sign, and the anti-chiral primary fields
for the $-$ sign.

It is now interesting to study the consequences of the twist
(2.17) on our $(b,c,\beta,\gamma)$ systems. We first analyze
the $+$ case.
{}From (2.10) and (2.11) we simply get
$${\hat T}_+ = T + \o{1}{2}\pa J = \sum_{i=1}^N \left(
\partial  c_i\, b_i + \gamma_i \, \partial
\beta_i \right) \ \ .
\eqno (2.19)
$$
This is the canonical stress-energy tensor
for a collection of $N$ commuting $(\beta,\gamma)$-systems
of  weight $\lam =0$, and $N$ anticommuting $(b,c)$-systems of
weight $\lam = 1$. ${\hat T}$ in (2.19) closes a Virasoro algebra
with vanishing central charge. Indeed, the central
charge of a first-order system of weight $\lambda$ is
$$
c_\lambda = \varepsilon  \left(1-3 Q^2\right)
\eqno (2.20)
$$
where
$$
Q = \varepsilon \left(1- 2\lam\right)
\eqno (2.21)
$$
is a ``background charge'' and $\varepsilon =1$ or $-1$, depending
on whether the
system is anticommuting or commuting.
In our case both the $(\beta,\gamma)$-systems
and the $(b,c)$-systems have $Q=-1$, but since their statistics
is different, their central charges exactly cancel.

To fully appreciate the effects of this topological twist
on our models, we now write the topological Lagrangian and
its BRST symmetries.
The Lagrangian is
$$
{\cal L}_{\rm top} = \sum_{i=1}^N
\left[ \gamma_i \bar \partial \beta_i +
\tilde\gamma_i  \partial \tilde \beta_i
- b_i \bar \partial c_i - \tilde b_i \partial \tilde c_i\right]
+ \sum_{i,j=1}^N \left[b_i  {\tilde b}_j \partial_i
\tilde\partial_j W \left (X\right )
\right]
\eqno(2.22)
$$
The BRST transformations which leave (2.22) invariant, can be
obtained from (2.8) and their analogues
by identifying the BRST parameter $\theta$ with
$\o{\epsilon^{+}}{\sqrt{2}}=\o{\bar\epsilon^{+}}{\sqrt{2}}$ (the
factor of $1/\sqrt{2}$ is introduced for convenience).
These transformations
are most conveniently
exhibited as the action of the
nilpotent Slavnov operator $s$  on all  fields,
namely
$$\eqalign{ s \, \beta_i & = 0    \ \ ,             \cr
 s \, {\tilde \beta}_i & = 0      \ \ ,             \cr
s \, b_i & =   \partial \beta_i    \ \ ,      \cr
s \, {\tilde b}_i & ={\bar \partial}{\tilde \beta}_i
     \ \ ,\cr
s \, \ga_i & = \partial c_i -
\sum_{j=1}^N b_j\,\partial_i \partial_j W \ \ ,\cr
s \, {\tilde \ga}_i & = {\bar \partial }{\tilde c}_i  +
\sum_{j=1}^N \tilde b_j\,\partial_i
\tilde \partial_j   W  \ \ ,\cr
s \, c_i &= +\partial_i W \ \ ,\cr
s \, {\tilde c}_i  &= - \tilde\partial_i W
\ \ .\cr}
\eqno(2.23)
$$
Using (2.23) it is quite easy to construct the representatives of the
BRST-cohomology classes and the corresponding integrated invariants.
According to the general theory,  we have to
consider multiplets
composed by a 0-form $\Phi_{P}^{ }$, a 1-form $\Phi_{P}^{(1)}$ and a
2-form $\Phi_{P}^{(2)}$
which satisfy the following descent equations
$$\eqalign{ s \, \Phi_{P}^{ } &=0\ \ ,\cr
s \, \Phi_{P}^{(1)} &=\,  -\, d \, \Phi_{P}^{ }\ \ ,\cr
s \, \Phi_{P}^{(2)} &=\,  -\, d \, \Phi_{P}^{(1)}\ \ ,\cr
d\, \Phi_{P}^{(2)} &= 0 \ \ .\cr}\eqno (2.24)$$
Moreover the 0-form $\Phi_{P}^{ }$ must belong to a non-trivial
BRST-cohomology class, {\it i.e.} it
should not be BRST-exact.
The solutions of the descent  equations (2.23)
provide the local physical observables $\Phi_{P}^{ }$
appearing
in  correlation functions as well as
the integrated invariants $\Phi_{P}^{(2)}$
which can be used to deform the theory.
Thus, the general form of a perturbed  topological
correlation function is
$$ c_{P_1,\dots,P_m}
(t_1,\dots,t_M)  = \left \langle \Phi^{ }_{P_1}(z_1) \,
\cdots \Phi^{ }_{P_m}(z_m) ~
\exp\left [\sum\limits_{k=1}^M  t_k \int\Phi^{(2)}_{P_k} \right ]\right
\rangle_{\rm top}
\eqno (2.25) $$
where  $\langle \cdots \rangle_{\rm top}$
means functional integration with the measure provided by the
umperturbed Lagrangian ${\cal L}_{\rm top}$
and $t_k$ are coupling constants
parametrizing its deformations $\int \Phi^{(2)}_{P_k}$.

In our $(b,c,\beta,\gamma)$ formulation
the general solution of the descent
equations (2.24) is
$$\eqalign{ \Phi_{P}^{ } &=   P(X)  \ \ ,
                                  \cr
 \Phi_{P}^{(1)} &= - \sum_{i=1}^N\left[ b_i\, \pa_i P  \, dz
 +\tilde b_i\,\tilde\pa_i P  \, d\bz \right]\ \ , \cr
\Phi_{P}^{(2)} &= \sum_{i,j=1}^N\left[b_i {\tilde b}_j \,
\pa_i\tilde\pa_j P \right] \, dz \wedge d \bz \ \ , \cr}
\eqno (2.26)$$
where $P(X)$ is any polynomial in the variables $X_i = \beta_i
{\tilde \beta}_i $ corresponding
to a non trivial element of the local ring determined
by the superpotential $W$ of
the Lagrangian (2.22). Indeed if the polynomial $P(X)$ is proportional to the
vanishing
relations ({\it i.e.} if
$P(X) =\sum_{i}
p^{i} (X) \o {\pa W}{\pa X_i}$), then using the BRST transformations
(2.23), we easily see that $P(X)  =  s \, K $ and so
$\Phi_{P}^{ }$ would be exact. (For the proof it suffices to
set $K = p^{i}(X) \,
\o {\pa \beta_j}{\pa X_i} \, c_j\ .$) Thus, the physical
observables in the topological theory are simply {\it local}
polynomials of $\beta_i$ and $\tilde \beta_i$, which
correspond to chiral primary fields of the original
N=2 superconformal theory.

On the other hand,  comparing the expression
of the 2-form  $\Phi^{(2)}_{P}$ in (2.26) with
the topological Lagrangian (2.22), it is easy to see that
a deformation of the potential
with some element $P(X)$ of the local ring, {\it i.e.}
$$
W(X) \longrightarrow W(X) -  t_P \, P(X) \ \ ,
\eqno(2.27)
$$
corresponds to a perturbation of the action
with $\int\Phi^{(2)}_{P}$, {\it i.e.}
$$
\int d^2z ~{\cal L}_{\rm top} \longrightarrow  \int d^2z~{\cal L}_{\rm top} -
t_P
\int\Phi^{(2)}_{P} \ \ .
\eqno(2.28)
$$
Thus, the possible perturbations of the theory are
in one-to-one correspondence with the possible
deformations of the potential.
As we are going to
see , something similar happens also in the
ordinary LG models, but only up to BRST-exact
terms.

For the sake of comparison we now write the general form of
the Lagrangian,
of the supersymmetry transformations and,
after twisting, of the topological BRST-transformations
of an ordinary N=2 LG model [33,34,41].
A short rheonomic derivation of the results hereinafter
reported is given in Appendix A.
Let $X^{i}(z,\bz)$ be $N$ complex scalar fields,  $X^{i^{\star}}(z,\bz)$ their
complex
conjugates,  $\psi^{i}$ and  $\psi^{i^{\star}}$ their  left-moving
anticommuting
superpartners, and ${\tilde \psi}^{i}$ and ${\tilde \psi}^{i^{\star}}$ their
right-moving  anticommuting superpartners. The
Lagrangian for a LG model with superpotential
$W$ is
$$\eqalign{ {\cal L}  = &- \left [ \pa \Xi \, \pab \Xjs + \pab \Xi \, \pa \Xjs
 \right ]\eta_{ij^*}
+ 8  \,\pa_i W \, \pa_{j^{\star}} {\bar W} \, \eta^{ij^{\star}} \cr
& +
4\,{\rm i} \left [ \posi \, \pab \posjs + \posib \,
\partial \posjsb  \right ] \eta_{ij^{\star}} \cr
& +  8 \left [ \pa_i \pa_j W \, \posi \posjb -
\pa_{i^{\star}} \pa_{j^{\star}} {\bar W} \, \posis \posjsb  \right ] }
\eqno (2.29)$$
where $\eta_{ij^{\star}}$ is the flat K\"ahlerian metric of ${\bf C}^n$.
Here we have understood summations over repeated indices,
and used the short-hand notations $\pa_i \equiv \pa/\pa \Xi$ and
$\pa_{i^\star}\equiv\pa/\pa \Xis$.
The  Lagrangian  above is invariant against the following  global N=2
supersymmetry
transformations
$$\eqalign{ \delta \Xi &= - \varepsilon^{-} \posi -
{\tilde \varepsilon}^{-} \posib   \ \ ,\cr
\delta \Xis &= + \varepsilon^{+} \posis +
{\tilde \varepsilon}^{+} \posisb \ \ ,\cr
\delta \posi & = - \o {\rm i}{2} \pa \Xi \, \varepsilon^{+} +
  \eta^{ij^{\star}}
\pa_{j^{\star}} {\bar W} \, {\tilde \varepsilon}^{-}\ \ ,\cr
\delta \posib  & = - \o {\rm i}{2} \pab \Xi \, {\tilde \varepsilon}^{+} -
\eta^{ij^{\star}}
\pa_{j^{\star}} {\bar W} \, \varepsilon^{-} \ \ ,\cr
\delta \posis & = \o {\rm i}{2} \pa \Xis \, \varepsilon^{-} +
 \eta^{ji^{\star}}
\pa_{j} W \, {\tilde \varepsilon}^{+} \ \ ,\cr
\delta \posisb & = \o {\rm i}{2} \pab \Xis \, {\tilde \varepsilon}^{-} -
\eta^{ji^{\star}}
\pa_{j} W \, \varepsilon^{+} \ \ .\cr}
\eqno(2.30)
$$

Contrary to our $(b,c,\beta,\gamma)$
formulation,  the
global supersymmetries (2.30) do not extend to classical superconformal
symmetries of the action (2.29), even
when the superpotential $W(X)$ is a quasi-homogeneous
function. Indeed, it is only after quantization
that one can argue
the equivalence of (2.29) at its infrared fixed point
with a (2,2) superconformal model.
Our theory in (2.2) instead, is superconformal already
at the classical level whenever the
potentials $V$ and ${\tilde V}$ are quasi homogeneous.
Of course, this applies in particular to the
left-right symmetric case we are discussing where we have a single potential
$W(\beta {\tilde \beta})$ that can be identified with the superpotential
$W(X)$ of the LG theory.

Performing the topological twist does not modify
the Lagrangian (2.29) but merely changes the spin of the
fields [41].  If we choose as BRST-parameter
$\theta={\varepsilon}^{+} ={\tilde \varepsilon}^{+}$
(as is appropriate for the $+$ twist),
the action of the topological Slavnov operator on the
LG fields turns out to be
$$\eqalign{ s\, \Xi  &=  0 \ \ ,\cr
  s \,\Xis &= \posis + \posisb \ \ ,\cr
  s \,\posi  &=- \o {\rm i}{2} \,\pa \Xi \ \ ,\cr
  s \,\posib  &=- \o{\rm i}{2}  \,\pab\Xi \ \ ,\cr
  s \,\posis  &= \eta^{i^{\star}j} \, \pa_j W \ \ ,\cr
  s \,\posisb &=  - \eta^{i^{\star}j} \, \pa_j W
\ \ .\cr}
\eqno(2.31)
$$

Using (2.31), we can easily solve the descent equations (2.23)
and find
$$
\eqalign{ \Phi_{P}^{ } &= P(X) \ \ ,\cr
\Phi_{P}^{(1)} &=  -2 \,{\rm i} \pa_i \,P
\left (\posi \, dz
+ \posib\, d{\bz} \right ) \ \ ,\cr
\Phi_{P}^{(2)} &=  - 4 \left [\pa_i \pa_j P\,
\posi \posjb \,+\, \pa_k P \, \pa_{l^{\star}} {\bar W} \, \eta^{kl^{\star}}
\right ] dz
\wedge  d \bz \ \ ,\cr}
\eqno(2.32)
$$
where $P(X)$ is a polynomial corresponding
to some non trivial element of the local
ring determined by the superpotential $W(X)$.
Indeed,  if $P(X)$ is proportional to the vanishing
relations ({\it i.e.} if $P(X)=\sum_ip^{i} (X) \o {\pa W}{\pa X_i}$), then
using
the BRST transformations
(2.31), one can  see that $P(X) = s \, K $ and
so $\Phi_{P}^{ }$ would be exact. (For the proof
it suffices to set $K = p^{i}(X)
\posjs \eta_{ij^{\star}}\ .$)

It is interesting to observe that under the
deformation
$$
W \longrightarrow  W - \o{1}{2} t_P  \, P(X) \ \ ,
\eqno (2.33)
$$
where $P(X)$ is some element of the local ring and $t_P$ is
the corresponding coupling constant,
the (topological) LG action changes as follows
$$\int d^2z ~ {\cal L} \longrightarrow  \int d^2z ~{\cal L} - t_P
\int \Phi^{(2)}_P  - {\bar t}_P \int{\bar \Phi}^{(2)}_{P}
\eqno(2.34)
$$
where ${\cal L}$ is given in (2.29), $\Phi^{(2)}_P$ in (2.32) and
${\bar \Phi}^{(2)}_{P}$ is the complex
conjugate 2-form. These equations have to
be compared with the analogous ones (2.26) and (2.27)
of the $(b,c,\beta,\gamma)$ theory.
At first sight, in the LG models there seem to be a
problem in identifying the topological pertubations
of the Lagrangian with the deformations of the superpotential
because of the last term in (2.34). However, this problem does
not exist because the 2-form
${\bar \Phi}^{(2)}_{ P}$ is BRST-exact, and so
adding or not its integral to the action is completely irrelevant.
In fact,  using  the BRST-transformations
(2.31), one can  check that
$$ {\bar \Phi}^{(2)}_{P}  = s \, \left ( - 4 \, \pa_{j^{\star}} {\bar P} \,
\posjs
\right )
\eqno(2.35)
$$
We want to emphasize that in the $(b,c,\beta,\gamma)$ formulation
instead, there is no counterpart
of this BRST-trivial
part and  deformations of the superpotential identically
coincide with topological deformations
of the Lagrangian.

We conclude this section by briefly commenting on the other choice of
sign in the topological twist for
our $(b,c,\beta,\gamma)$-system. If one chooses in (2.17) the $-$ sign,
from (2.10) and (2.11) one obtains
$$\eqalign{
{\hat T}_{-}&= T -\o{1}{2}\pa J\cr
&= \sum_{i=1}^N(1-2\lam_i)\,\partial  c_i \, b_i  - 2\lam_i \, b_i \partial c_i
+  (1-2\lam_i) \,\gamma_i\, \partial \beta_i  -  2\lam_i \,  \beta_i \,
\partial \gamma_i \ \ .}
\eqno (2.36)
$$
This is the canonical stress-energy tensor
for $N$ commuting $(\beta,\gamma)$-systems with weight $2\lambda_i$
and $N$ anticommuting $(b,c)$-systems also with weight $2\lam_i$.
It is straighforward to check that $T_-$ closes a Virasoro
algebra with zero central charge; indeed the bosonic and fermionic
contributions to the central charge exactly cancel each other.
However, the cohomology classes of the BRST charge $Q_-^{\rm BRST}$  correspond
to anti-chiral primary fields of the original N=2 algebra and these do not
have a simple and local
representation in terms of the elementary fields appearing in
the Lagrangian: indeed, to describe the anti-chiral operators one has
to resort to the bosonization of the $(b,c,\beta,\gamma)$-systems (see
Section 4).
On the other hand, as we explicitly show in appendix B, after
performing the tological twist,
the Lagrangian is BRST-exact,
{\it i.e.} it is of the form $ {\cal L}_{\rm top}
= \ncomm {Q^{\rm BRST}_-}{ {\cal L}'}$ for some local
functional  ${\cal L}'$.
Using the terminology of [22], this means that the $-$ twist defines
a topological field theory of the Witten-type.  On the contrary, the $+$ twist
leads to the Lagrangian (2.22) which is not BRST exact with respect
to $Q_+^{\rm BRST}$; thus the $+$ twist defines
 a topological field theory of the Schwarz-type.
As pointed out in [41], also the ordinary topological LG models are
theories of the
Schwarz-type.

In conclusion, we have shown that N=2 LG models
admit a $(b,c,\beta,\gamma)$-formulation  which is already superconformal at
the
classical level. After topological twisting, there is a natural correspondence
between the deformations of the LG potential and
the  abstract topological deformations. In the next sections, after discussing
the
renormalization group properties of our theory, we shall illustrate how one can
use this explicit formulation to calculate (perturbed) topological correlation
functions in LG models.
\vskip 2truecm

\centerline{\bf 3. On the quantum properties of the $(b,c,\beta,\gamma)$
system}
\vskip 0.4truecm
In the previous section we discussed the classical properties of the
action (2.1) and showed that with a suitable choice of
the interaction potential, the theory
exhibits a non trivial (2,2)-superconformal invariance.
However, the presence of interactions
can in principle spoil this invariance at the quantum level
and one has eventually to restore it after a suitable
renormalization [42].
In this section we are going to show that
no loop corrections are present in our model
so that the classical results automatically extend to the quantum theory.

For the sake of clarity, we begin by considering a single left-right
symmetric $(b,c,\beta,\gamma)$-system of weight $\lambda=\tilde \lambda$
with potential
$$
W= \o{1}{n+1} (\beta \tilde \beta)^{n+1} \ \ .
\eqno(3.1)
$$
This corresponds to the $A_{n}$ minimal model of the N=2 discrete
series if $\lambda = 1/(2n+2)$ (see (2.13)). However, for the time being,
we leave
the weight $\lambda$ unfixed.
The Lagrangian for this system is ${\cal L} = {\cal L}_0 + \Delta {\cal L}$
where ${\cal L}_0$ is as in (2.2a) and the interaction term is
$$
\Delta {\cal L}=g ~b \tilde b \,\beta^n \tilde \beta^n\eqno(3.2)
$$
where $g$ is a coupling constant. Since the weight of $\beta$ and $\tilde
\beta$
is arbitrary, $g$ is a quantity with dimension
$$
[g]=(1- 2\lam(n+1)) \ \ .
\eqno(3.3)
$$
To study the scaling properties of this system, we compute
the trace
of the stress-energy tensor which turns out to be
\footnote{$^3$}{Here and in the following ``c.c.'' means
exchanging the untilded fields with the tilded ones and $\part$ with $
\bar \part$.}
$$
T_{z\bar z}= \left[-\lambda \beta \bar\partial \gamma + (1- \lam)\gamma
\bar\partial \beta -(\lambda +\um)b \bar\partial c
-(\um-\lambda)c \bar\part b  +
g ~b \tilde b \beta^n \tilde \beta^n\right ] + \hbox{c.c.}\ \ .\eqno(3.4)
$$
After using the equations of motion (2.3a), we have
$$
\Theta \equiv -T_{z \bar z}=g (2(n+1)\lambda - 1)\,b \tilde b \,\beta^n
\tilde \beta^n \ \ ,
\eqno(3.5)
$$
so that our system is classically invariant under scale transformations
({\it i.e.} $\Theta =0$) either if
$$
g =0 \quad \hbox{~ for any} \quad \lambda \ \ ,
\eqno(3.6{\rm a})
$$
or if
$$
\lambda =\o{1}{2(n+1)} \quad \hbox{~ for ~} g\ne 0 \ \ .
\eqno(3.6{\rm b})$$
Discarding the case (3.6a) which corresponds to a free theory, we see
from (3.6b) that $\lambda$ must be fixed by the homogeneous weight of the
potential (cf. (2.7)); when (3.6b) is satisfied of course $g$
becomes dimensionless and the operator $b\tilde b\,\beta \tilde \beta$ becomes
marginal, so that no dimensionful parameters are left in the model.

Let us now quantize this system by using perturbation theory in $g$.
{}From the explicit expression of the Lagrangian ${\cal L}$, we see
that the propagators are
$$
\eqalign{
\langle \gamma(z,\bar z) \beta (w, \bar w) \rangle &=
\langle b(z,\bar z) c (w, \bar w) \rangle =\o{1}{z-w} \ \ ,\cr
\langle \tilde\gamma(z,\bar z) \tilde\beta (w, \bar w) \rangle
&=\langle \tilde b(z,\bar z) \tilde c (w, \bar w) \rangle
=\o{1}{\bar z-\bar w} \ \ , \cr}\eqno(3.7)
$$
so that it is obvious that even when
the interaction (3.2) is present, it is impossible to form loops.
Therefore we conclude that there are no (perturbative) quantum corrections to
the classical results
simply because there are no loops!
These considerations imply in particular that
$\Theta$ in (3.5) is also the {\it quantum} trace of the stress-energy tensor
and hence the coefficient of the spinless operator $b \tilde b \beta^n \tilde
\beta^n$
appearing in (3.5)
can be interpreted as a renormalization
group $\beta$--function [43], namely
$$
\beta (g) = g\,(2(n+1)\lambda -1) \ \ .
\eqno(3.8)
$$
The zeroes of $\beta (g)$ identify the conformal fixed points and these are
given
precisely by (3.6a) and (3.6b).

It is now interesting to see what happens when a second interaction
$$
\Delta{\cal L}^\prime=
g^{\prime}b \bar b \beta^m \bar \beta^m ~~~\quad (m < n)\eqno(3.9)
$$
is added to the original system.
We now assume that $\lambda= 1/(2n+2)$ so
that (3.9) can be considered as a
perturbation around a conformal theory. Following the same procedure as
above, we compute the $\beta$--function $\beta(g')$ and find
$$
\beta(g^\prime)= g^{\prime} \left(\o{m+1}{n+1} - 1\right)=\o{m-n}{n+1} g^\prime
\ \ .
\eqno(3.10)
$$
It is clear from (3.10) that the
new model does not have any
non-trivial fixed point; indeed the only solution to $\beta(g^\prime)=0$ is
$g^\prime=0$ which is achieved in the ultraviolet regime
for $m<n$. Hence
we cannot have a renormalization
group flow to another N=2 superconformal
field theory, in agreement with the conclusions of [44].

The extension of these results to the generic case of quasi-homogeneous
potentials is an easy task. To this end let us first recall that if
$f$ is a quasi--homogeneous polynomial in $N$ variables
with weights $(\omega_1, \cdots \omega_N)$, ($\omega_i \in Q$,
$\omega_i >0$) and
$$f=\sum_\rho a_\rho x^\rho \eqno(3.11{\rm a})$$
where $\rho\equiv(\rho_1,\cdots \rho_N)$, $X^\rho\equiv X_1^{\rho_1} \cdots
X_N^{\rho_N}$, $\rho_i \in Z^+$ and $a_\rho \ne 0$, then
$$
\rho_1 \omega_1 + \cdots \rho_N \omega_N=1\ \ . \eqno(3.11{\rm b})
$$
Let us now consider the following interaction
term
$$
\Delta {\cal L}=g \sum_{i,j}\bit \bjt \part_i V \tilde \part_j \tilde V
\eqno(3.12)
$$
where $V(\beta)$ and $\tilde V(\tilde \beta)$ are quasi--homogeneous
potentials satisfying (3.11).
For simplicity, we take $V(\beta)=\tilde V(\tilde\beta)$ and assume that the
weights
$\lambda_i=\tilde \lam_i$
are unconstrained. Then, the
trace of the stress--energy tensor, upon using the equations of
motion (2.3a), turns out to be
$$\eqalign{
\Theta = - T_{z\bar z}= &-g \sum_{i,j}\left(\bit \bjt \part_i V \tilde
\part_j \tilde V -2\lam_i\bit \bjt \part_i V \tilde
\part_j \tilde V \right) \cr
&-\sum_{i,j,l}\left(\lambda_i \bei b_l \bjt \part_i \part_l V
\tilde \part_j \tilde V
- \lambda_i \beit b_j \tilde b_l \tilde\part_i \tilde\part_l \tilde V
\part_j  V \right) \ \ . \cr} \eqno(3.13)
$$
Using (3.11), after some algebra, the trace (3.13) can be rewritten as
$$
\Theta=g \left(2 \sum_k \lambda_k \rho_k - 1\right)\left(
\sum_{i, \rho}a_\rho b_i \rho_i \beta_1^{\rho_1}\cdots
\beta_N^{\rho_N}\right)\left(
\sum_{j, \rho}\tilde a_\rho \tilde b_j \rho_j \tilde{\beta}_1^{\rho_1}\cdots
\tilde{\beta}_N^{\rho_N}\right)\ \ .\eqno(3.14)
$$
If $g\not =0$, the system is invariant under scale transformations only if
$$
\sum_i \lambda_i \rho_i=\o{1}{2}\eqno(3.15)
$$
Comparing (3.15) with (3.11) we see that the weights
$\lam_i$ must be one half of the
homogeneous weights of the potential $\omega_i$. Since there are no loop
corrections, this result extends automatically to the quantum theory.
Furthermore, we point out that the same conclusion is obtained in a similar way
when $V(\beta)$
and $\tilde V (\tilde \beta)$ are different, or when the interaction depends
on a single quasi--homogeneous function $W$ in the variables
$X_i=\beta_i\tilde\beta_i$
with weight $\omega_i$.

Finally, in our formulation it is easy to realize that the potential
$$
\hat V (\beta^{(i,A)})=V(\beta^i) + \sum_{A=n+1}^{m+n} (\beta^A)^2
\eqno(3.16)
$$
defines the same conformal theory as the potential $V$
(as one should expect from the notion of stable
singularity [35,36]).
Indeed, a $(b,c,\beta,\gamma)$-system with $\lambda_A=\o{1}{4}$
gives a $c=0$ conformal field theory.
\vskip 2truecm

\centerline {\bf 4.  Bosonization of the $(b,c,\beta,\gamma)$-system }
\centerline{\bf        before and after topological twisting}
\vskip 0.3cm
The result of our previous analysis is that for any value $
\lambda=(2n+2)^{-1} $
with $ n=2,3,\dots$, we have a realization of the N=2 superconformal algebra
with central charge
$$ c= 3-12\lam = \o {3(n-1)}{n+1} \ \ .
\eqno(4.1)$$
The conformal weights $h$ and the $U(1)$-charges $q$
of the pseudo-ghost fields that define such a realization
are given by
$$\eqalignno{
h(\beta) & = \o{1}{2(n+1)}  \; \; \; \;  , \; \; \; \; \hfill
 q(\beta)  =  \o{1}{n+1} \ \ ; &(4.2{\rm a})\cr
h(\gamma) & = \o{2n+1}{2(n+1)}  \; \; \; \;  , \; \; \; \; \hfill
 q(\gamma) =  -  \o{1}{n+1} \ \ ; &(4.2{\rm b})\cr
h(b) & = \o{n+2}{2(n+1)}  \; \; \; \;  , \; \; \; \; \hfill
 q(b) =  -  \o{n}{n+1} \ \ ;&(4.2{\rm c})\cr
h(c) & = \o{n}{2(n+1)}  \; \; \; \;  , \; \; \; \; \hfill
 q(c)  =  \o{n}{n+1} \ \ . &(4.2{\rm d})\cr
 }$$
The Fock space generated by the modes of the almost-free fields $b$,
$c$, $\beta$, $\gamma$
and their spin fields, contains the irreducible representations
of the N=2 minimal models. Such representations
can be obtained from the Fock space
through a suitable projection like in the case of the
standard free-field realization of the minimal models as
given for instance in [38].
In this section our aim is
to make contact with this Coulomb gas formalism, which, as we will see
in the sequel, enables
us to calculate explicitly (perturbed) topological
correlation functions in the presence of a LG interaction.

Adopting the conventions of [38], an N=2 minimal model with central charge
as in (4.1), can be described in terms of three scalar fields
$\phi_0$, $\phi_1$ and $\phi_2$ with mode expansions
$$
\phi_i(z) = {\hat q}_i- {\hat p}_i \ln z + \sum_{k\not = 0}
\o{{\hat a}^i_k}{k} z^{-k} \ \ \ , \ \ \ i=0,1,2 \ \ \ ,
\eqno(4.3)
$$
where
$$
[{\hat q}_i, {\hat p}_j] = \delta_{ij} \ \ , \ \
[{\hat a}^i_k,{\hat a}^j_\ell]
= k \, \delta_{k+\ell,0}\ \delta^{ij} \ \ .
\eqno(4.4)
$$
While $\phi_0$ and $\phi_2$ are really free fields, $\phi_1$
is coupled to a background charge
$$Q_1^{(n)} = \sqrt{\o{2}{n+1}} \ \ .\eqno(4.5)$$
In this realization the holomorphic currents of the N=2 superconformal
algebra are \footnote{$^4$}{Here and in the following, any exponential of free
fields is understood as normal ordered.}
$$\bosoncurrentn2 {4.6}$$
The field $\phi_0$ bosonizes the U(1)-current and its exponentials
realize the
well known N=2 spectral flow [5]. The operators $\Psi^{\pm}$ in (4.6d)
are,  instead, parafermionic currents and generate the non trivial part of the
N=2 algebra.
The complete Fock space which embeds the N=2 irreducible modules is generated
by
the vertex operators
$$V_{q,\ell,m} = \vertenne{q}{\ell}{m}
\eqno(4.7)$$
and their derivatives.

In particular the N=2 primary fields are  given by
$$
\Lambda_{\ell,m;s}^{(n)}= \o{1}{\sqrt{2}}
\vertenne{m+sn-s}{\ell}{m}
\eqno(4.8)
$$
where $\ell$ takes the integer values  $0 \le \ell \le n-1$ and
$m$ takes the integer values $ m = -l,-l+2,...,l$.
The quantum number $s$ represents the sector and is
$0$ in the Neveu-Schwarz sector and $\pm 1/2$ in the Ramond sector.
The conformal weight $h$ and the $U(1)$-charge $q$ of $
\Lambda_{\ell,m;s}^{(n)}$ are given by the standard formulas
$$\eqalign{h(\ell,m;s) &= \o{\ell(\ell+2)}{4(n+1)} - \o{m^2}{4n-4} +
\o {(m+sn-s)^2}{2(n^2-1)} \ \ ,
\cr
q(m;s) & =  \o{m+sn-s}{n+1} \ \ .}\eqno(4.9) $$
As we will see later, it is convenient to factor out the $\phi_0$
contribution and rewrite the primary fields (4.8) as
$$
\Lambda_{\ell,m;s}^{(n)} = \exp\left[\sqrt{\o{3}{c}} q(m,s)\ \phi_0
\right] \ \varphi_m^\ell
\eqno(4.10)
$$
where
$$\parafer {\ell}{m} = \paraferbos {\ell}{m} \ \ .\eqno(4.11)$$
The operators $\parafer {\ell}{m}$ are the principal
primary fields of the ${\bf Z}_{n-1}$ parafermion algebra and must
be identified according to
$$
\parafer {\ell}{m} \sim \parafer {n-1-\ell}{m\pm(n-1)}\ \ .
\eqno(4.12)
$$
In fact the Hilbert space created by $\parafer {\ell}{m}$ is isomorphic
to the one created by $\parafer {n-1-\ell}{m\pm(n-1)}$ due to the
existence of a map between the two that commutes  with all generators
of the algebra [38].

In order to relate this realization of the N=2 minimal models to the one
provided by our
$(b,c,\beta,\gamma)$-system, we bosonize the latter according to the
standard
rules and write\footnote{$^5$}{Notice that the bosonization rules we are
giving
are actually true for the ``free'' holomorphic part of the $c, \gamma$
fields. However, as we are going to see, we only need of the bosonized
$b,\beta$ fields (which are correctly expressed by (4.13) (4.19a) and
(4.19b)) in application to topological correlation functions}
$$ \bcbetaboson \eqno(4.13)$$
where the $\pi_i$'s are scalar fields coupled to the following
background charges
$$
{\tilde Q}_1^{(n)} = -\o{1}{n+1} \ \ , \ \
{\tilde Q}_2^{(n)} = {\rm i}\o{n}{n+1} \ \ , \ \
{\tilde Q}_3^{(n)} = -1 \ \ .
\eqno(4.14)$$
These numbers are explained as follows: $\pi_1$ bosonizes the
anticommuting $(b,c)$-system whose weight is
$\lam + \um=-\o{1}{n+1}$.
Insertion of this value in the general formula (2.21)
yields ${\tilde Q}_1^{(n)}$ as listed in (4.14). The field
${\rm i}\pi_2$ bosonizes the commuting $(\beta,\gamma)$-system according to
the rule
$$\gamma = {\rm e}^{-{\rm i} \, \pi_2} \, \pa \xi \ \ , \ \
\beta = {\rm e}^{{\rm i} \, \pi_2} \, \eta \eqno(4.15)$$
where $\xi$ and $\eta$ form an anticommuting first-order system of
weight $\lam_{\xi\eta} = 1$. The background charge
${\tilde Q}_2^{(n)}$ of the field $\pi_2$ follows from (2.21)
with $\lam_{\beta\gamma}=\lam =\o{1}{2n+2}$.
Finally $\pi_3$ is the scalar field that bosonizes the
$(\xi,\eta)$-system and its backround charge ${\tilde Q}_3^{(n)}$
also follows from (2.21) upon use of the value $\lam_{\xi\eta}=1$.

Consequentely, in terms
of the fields $\pi_i$'s , the stress-energy tensor of the N=2 model
is
$$\stresspi \eqno(4.16)$$
Similarly, using (4.13) in (2.9) and (2.11) we obtain
$$\eqalignno{J & = \o{1}{n+1}\left(n\pa \pi_1 -{\rm i}\pa
\pi_2 \right)\ \ ,&(4.17{\rm a})\cr
G^{-} &= 2\sqrt{2} \exp \left[- \pi_1 -{\rm i} \, \pi_2  +
\pi_3\right] \pa \pi_3 \ \ .&(4.17{\rm b})\cr}$$
Comparing (4.16) and (4.17) with (4.6), we obtain the
relation between the $\pi$'s and the $\phi$'s, namely
$$\conversion {4.18}$$

One can proceed even further and use (4.13) and (4.18) to identify the
pseudo-ghost fields with the operators of the abstract
N=2 superconformal model. Expicitly one finds
$$\eqalignno{\beta &= \vertenne{1}{1}{1}\ \ , &(4.19{\rm a})\cr
b  &= \vertenne{-n}{1}{-1}\ \ , &(4.19{\rm b})\cr
c &= \vertenne{n}{-1}{1} \ \ , &(4.19{\rm c})\cr
\gamma &= \vertenne{-1}{-1}{-1} \times \cr
& \ ~ ~ \  \times \ \sqrt{\o{n-1}{2}}  \left (- \sqrt{\o{n+1}{n-1}} \pa
\phi_1 + {\rm i}\pa\phi_2 \right ) \ \ . &(4.19{\rm d})\cr}
$$
{}From (4.19a) one realizes that $\beta$ is a chiral primary field
and is given by
$$
\beta=\Lambda_{1,1;0}^{(n)} \ \ .\eqno(4.20{\rm a})
$$
More generally one can write
$$\beta^\ell=\Lambda_{\ell,\ell;0}^{(n)} \ ~ ~ \ \hbox{for \ }
\ell=0,\dots,n-1\ \ ,
\eqno(4.20{\rm b})$$
which shows that at the quantum level the general
chiral primary field is simply the $\ell$-th  power of $\beta$ and the
vanishing relation is recovered by enforcing
the bound $\ell\le n-1$. Moreover $b$
is the first component of a chiral primary superfield
and can be explicitly obtained by $\o{1}{2\sqrt{2}}\oint_z G^-(w)
\beta(w)=b(z)$; the same is true for fields of the form
$b\beta^{l-1}$.
On the contrary $\gamma$ and $c$ are in the Fock space of the
three scalar fields, but not in
the N=2 irreducible module.

We now consider the case when the theory
is topologically twisted with
$$
Q_+(z) = G^+(z) \ \ .\eqno(4.21)
$$
As mentioned in Section 2, we have a new $(b,c,\beta,\gamma)$-system with
$\lam_{
\beta}=0$, $\lam_b=1$  whose Lagrangian is given in (2.22). These new
pseudo-ghost
fields are still bosonized as in (4.13), but now the background charges
of the $\pi_i$'s become
$$
{\tilde Q}_1^{(n)} = -1 \ \ , \ \
{\tilde Q}_2^{(n)} = {\rm i} \ \ , \ \
{\tilde Q}_3^{(n)} = -1 \ \ .
\eqno(4.22)$$
The new stress-energy tensor is then given by
$$
\hat T=\o{1}{2}\left [ (\part \pi_1)^2 + (\part \pi_2)^2 +
(\part \pi_3)^2 \right ]-\o{1}{2} \left (-\part^2 \pi_1 +{\rm i}
\ \part^2 \pi_2
-\part^2 \pi_3 \right) \ \ , \eqno(4.23)
$$
whereas the $U(1)$-current $J$ is the same as in (4.17a).
On the other hand, the twist of the stress-energy tensor
is given by $\hat T=T+\o{1}{2}J$, and hence, using (4.6a,b) we get
$$
\hat T=\o{1}{2}\left [ (\part \phi_0)^2 + (\part \phi_1)^2 +
(\part \phi_2)^2 \right ]-\o{1}{2} \sqrt{\o{2}{n+1}}
\part^2 \phi_1 +\o{1}{2} \sqrt{\o{n-1}{n+1}}\part^2 \phi_0  \ \ .
\eqno(4.24)
$$
If we now compare (4.23) and (4.24) and observe that $J$ is the same before
and after the twist, so that
$$
J=\sqrt{\o{n-1}{n+1}}\part \phi_0=\o{n}{n+1} \part \pi_1 -\o{i}{n+1}
\part \pi_2 \ \ ,\eqno(4.25)
$$
we can realize that the relations (4.18) and the identifications
(4.19) and (4.20) hold
true also in the topological field theory, giving us a complete
characterization of the fields $b$, $c$, $\beta$ and $\gamma$
at the quantum level.
Moreover, using
(4.19)  and taking into account both left
and right movers, we can easily check
the descent equations (2.24) in the complete bosonized formalism.
As is clear from (4.24) in comparison with
(4.6a), the net effect of the topological
twist is simply to switch on
a background charge for $\phi_0$ given by
$$
Q_0^{(n)} = \o{1-n}{\sqrt{n^2-1}} \ \ .
\eqno(4.26)
$$
Therefore, even if the bosonized expressions for the topological
$b$, $c$, $\beta$ and $\gamma$ are still given
by (4.19), their conformal dimensions
change with  the twist. In particular
the chiral primary fields $\beta^\ell$ loose their
conformal weight and become dimensionless, as is appropriate
for the physical operators of a topological field
theory. Furthermore, the $U(1)$ current $J$ aquires an anomaly
proportional to $Q_0^{(n)}$.

Let us  briefly  mention that if we perform the topological
twist with $Q_-=G^-$ instead of $Q_+=G^+$,
not only the the stress energy tensor but
also the lagrangian becomes a BRST commutator (see Appendix B for
details). In this case, however, there
is no identification of the chiral fields in terms of
{\it local} expressions of $b$, $c$, $\beta$ or $\gamma$:
they can only be written in a
bosonized form. As is well known, chiral fields with respect to
the  BRST charge $Q_-= G^-$ are
antichiral fields with respect to the BRST charge $Q_+=G^+$.
This means that the bosonized expression of these
fields could be obtained via spectral flow from (4.20).

The complete bosonization of the $(b,c,\beta, \gamma)$-system
we have just presented is the technical tool which
enables us to make explicit calculations
of (perturbed) correlation functions for single minimal models
as well as for tensor
products thereof. It is also very useful in establishing the
precise relationship between
the correlation
functions of topological
minimal models and the chiral Green functions of LG
theories as computed in [34].

To this end, let us first introduce the following notation
$$
|q,\ell, m\rangle \equiv \lim_{z\to 0} V_{q,\ell,m}(z)~|0,0,0\rangle
\eqno(4.27)
$$
where $V_{q,\ell,m}$ is defined in (4.7) and
$|0,0,0\rangle$ is the $Sl(2,{\bf C})$ invariant
vacuum of $\phi_0$, $\phi_1$ and $\phi_2$. Before the topological
twist only $\phi_1$ has a background charge and
the dual conjugate of $|q,\ell,m\rangle$
is $\langle -q,-2-\ell,-m|$. After the twist also $\phi_0$ acquires a
background charge and so the dual conjugate of $|q,\ell,m\rangle$
becomes $\langle n-1-q,-2-\ell,-m|$.

In the LG theory with superpotential $W\sim X^{n+1}$, the supersymmetric
vacua $|m\rangle$ ($m=0,\dots,n-1$)
are identified at the conformal point
with the Ramond vacua of the minimal model $A_{n}$.
According to (4.8), the Ramond chiral primary fields of such a model
are
$$
R_m(z)= {\cal N}_m \ \Lambda_{m,m;-\o{1}{2}}^{(n)}(z)
\eqno(4.28)
$$
where $m=0,\dots,n-1$ and  the normalization factor ${\cal N}_m$
is introduced to
enforce the standard structure constants of the N=2
operator algebra \footnote{$^6$}{Notice
that the Ramond fields $R_m$ do not have a local expression
in terms of the fields $b,c,\beta,\gamma$  of our model,
contrary to the Neveu-Schwarz
chiral primaries which are simply powers of the bosonic field $\beta$.}.
This normalization can be computed using different techniques
[34,39] and is given by
$$
{{\cal N}_m}^2= {1\over{(2n+2)^{m\over{n+1}}}}
\sqrt{{\sin\left(\pi\over{n+1}\right)}
\over{\sin\left({{\pi(m+1)}\over{n+1}}\right)}}
{{\Gamma\left(1\over{n+1}\right)}\over{\Gamma\left({m+1}\over{n+1}\right)}}
\ \ .\eqno(4.29)
$$
Therefore at the conformal point,
the supersymmetric vacua of the LG theory are
$$
\eqalign{
|m\rangle &= \lim_{z\to 0}R_m(z)|0,0,0\rangle_{\rm L}
\times \lim_{{\bar z}\to 0}
R_m({\bar z})|0,0,0\rangle_{\rm R} \cr
&= {{\cal N}_m}^2 |m-(n-1)/2,m,m\rangle_{\rm L} \times
|m-(n-1)/2,m,m\rangle_{\rm R} }\eqno(4.30)
$$
where the subscripts L and R refer to the holomorphic
and anti-holomorphic
components respectively.  The vacua $\langle m|$ are
obtained by taking the
dual conjugate of (4.30) and remembering that only
$\phi_1$ has a background
charge (indeed the topological twist has not been
performed yet).

It is now straightforward to compute the correlation
functions of a string of chiral primary
fields between two supersymmetric vacua.
{}From Eqs. (4.20b) and (4.30) we have
$$
\langle m_1| \prod_{i=1}^N \left( \beta(z_i)
{\tilde{\beta}}(\bar{z}_i) \right)^{\ell_i} |m_2\rangle
= {{{\cal N}_{m_2}}^2 \over {{\cal N}_{m_1}}^2}
\prod_{i=1}^N
(z_i \bar{z}_i)^{-{\ell_i}\over{2n+2}} \ \delta\left({\sum_i \ell_i +m_2-m_1}
\right)
\eqno(4.31)
$$
where the $\delta$-function arises from charge conservation.
Apart from the $z$-dependent factor, this
result coincides with the LG chiral Green functions
computed in [34] using
quantum field theory techniques.
To make the precise comparison, however, one has to remember that in
[34] the LG
theory was defined on a cylinder, whereas our formula (4.31)
applies to the plane.
This difference is easily eliminated by mapping the plane to the cylinder,
under which the chiral primary fields trasform as
$$
\beta^{\ell_i}(z_i) \longrightarrow \beta^{\ell_i}(w_i) \
z_i^{{\ell_i}\over{2n+2}} \ \ .
\eqno(4.32)
$$
Here $w_i$ are the cylinder coordinates. An
analogous expression holds also for the $\tilde{\beta}$ fields.
The $z$-dependent
factors of (4.31) are therefore cancelled in going from
the plane to the cylinder and we can conclude that
$$
\left. \langle m_1| \prod_{i=1}^N \left(\beta {\tilde \beta}\right)^{\ell_i}
|m_2\rangle
\right|_{\rm cyl} = ~ {{{\cal N}_{m_2}}^2 \over {{\cal N}_{m_1}}^2}
\ \ \delta\left({\sum_i \ell_i +m_2-m_1}
\right)
\eqno(4.33)
$$
exactly coincides (normalization factors included) with the
chiral Green function of
the LG theory of [34]. Once more we see that the L.G. field $X$
has to be
identified with the product $(\beta {\tilde \beta})$ (see also Eq. (2.16)).

We now proceed to establish the relationship with the topological
conformal field theories.
To this end let us consider a particular case of (4.31),
namely the correlation functions between the
lowest vacuum $|0\rangle$
and the highest one $\langle n-1|$,
$$
\eqalign{
& ~\langle n-1| \prod_{i=1}^N \left(\beta(z_i) {\tilde \beta({\bar
z}_i})\right)^{\ell_i}|0\rangle
{}~ = \cr
& {{\cal N}_0\over{\cal N}_{n-1}}
{}_{\rm L}\langle (1-n)/2,-1-n,1-n| \prod_{i=1}^N \beta^{\ell_i}(z_i)
|(1-n)/2,0,0\rangle_{\rm L} \times {\rm (c.c.)}\ \ .}
\eqno(4.34)
$$
Since
$$
|(1-n)/2,0,0\rangle =
\exp\left[\o{(1-n)/2}{\sqrt{n^2-1}} {\hat q}_0\right]~|0,0,0\rangle
\eqno(4.35)
$$
and
$$
\beta^\ell(z)~\exp\left[\o{(1-n)/2}{\sqrt{n^2-1}} {\hat q}_0\right] =
\exp\left[\o{(1-n)/2}{\sqrt{n^2-1}} {\hat q}_0\right]~\beta^\ell(z)
{}~z^{-{\ell}
\over{2n+2}}\ \ ,\eqno(4.36)
$$
from (4.34) and (4.31), we get
$$
{}_{\rm L}\langle 1-n,-1-n,1-n| \prod_{i=1}^N \beta^{\ell_i}(z_i)
|0,0,0\rangle_{\rm L} \times {\rm (c.c.)} =
 \delta\left(\sum_i \ell_i -n+1\right) \ \ .
\eqno(4.37)
$$
This is the natural candidate for a topological
correlation function. Notice that (4.37) is independent
of $z_i$
as any topological correlator should be.
Indeed all the $z$-dependent factors are canceled
in flowing from the lowest vacuum $|0\rangle$ of the Ramond sector
to the $Sl(2,{\bf C})$ invariant vacuum $|0,0,0\rangle$
of the Neveu-Schwarz sector.
However, the fields $\phi_0$, $\phi_1$ and $\phi_2$ which
implicitly appear in (4.37) are those which bosonize the original
$(b,c,\beta,\gamma)$-system {\it before} the topological
twist. To obtain a more adequate
characterization of topological
correlation functions in the bosonized formalism, it is more
appropriate to use the fields which bosonize a twisted
$(b,c,\beta,\gamma)$-system.
As we have seen, only very few things change; most notably the
field $\phi_0$ acquires the background charge (4.26).
Thus, in (4.37) instead of the state
$\langle 1-n,-1-n,1-n|$, which is the dual conjugate
of $|n-1,n-1,n-1\rangle$ before the twist, we should have the state
$\langle 0,-1-n,1-n|$ which is the conjugate of
$|n-1,n-1,n-1\rangle$ after the twist. If
we define the topological vacuum $|0\rangle_{\rm top}$ as
$$
|0\rangle_{\rm top} \equiv |0,0,0\rangle
\eqno(4.38{\rm a})
$$
and its dual conjugate ${}_{\rm top}\langle \infty|$ as
$$
{}_{\rm top}\langle \infty| \equiv \langle n-1,-2,0| \ \ ,
\eqno(4.38{\rm b})
$$
we can rewrite the holomorphic part of (4.37) and define
the topological correlator as follows
$$
\eqalign{
\langle \prod_{i=1}^N \beta^{\ell_i}(z_i) \rangle_{{\rm top}}  &
\equiv \langle 0,-1-n,1-n| \prod_{i=1}^N \beta^{\ell_i}(z_i) |0,0,0\rangle
\cr
& ={}_{\rm top}\langle\infty| \Omega^\dagger
\prod_{i=1}^N \beta^\ell_i(z_i) |0\rangle_{\rm top} \cr
& = \delta\left(\sum_i \ell_i-n+1\right)}
\eqno(4.39)
$$
where
$$
\Omega = \exp\left[
\o{n-1}{\sqrt{n^2-1}} {\hat q}_0
+\o{n-1}{\sqrt{2(n+1)}}{\hat q}_1 + {\rm i}\o{n-1}{\sqrt{2(n-1)}}{\hat q}_2
\right]
\eqno(4.40)
$$
and the $^\dagger$ operation is defined as $\left( {\rm e}^{\alpha
{\hat q_i}} \right)^\dagger
= {\rm e}^{-\alpha {\hat q_i}}$ (see e.g. [38]). Notice that $\Omega$
is simply the zero-mode part (the only one which survives
on the left vacuum) of the top
chiral primary field $\beta^{n-1}$.

Therefore our conclusion is that in the bosonized
formalism a topological correlation function of a string
of fields is obtained by taking the expectation
value between the topological
vacua $|0,0,0\rangle$ and $\langle 0,-1-n,1-n|$.
\vskip 2truecm

\centerline {\bf 5. Explicit calculations of topological correlation functions}
\vskip 0.3cm

In this section we show in a few examples how to use
the $(b,c,\beta,\gamma)$
representation to compute {\it explicitly} some
(perturbed) topological
correlation functions. We also verify that the parameters of
the deformed LG potential for the $(b,c,\beta,\gamma)$-system
are the flat coordinates of the topological field theories.
We stress that our techniques can be applied
to single minimal models as well as to their tensor products,
since there is practically no difference between the two
cases. Even though the final goal is to use our methods
in the interesting case of the Calabi-Yau 3-fold,
for the sake of clarity here we will limit ourselves
to the simpler cases of the minimal models and the torus.

We start by considering the simplest possible
situation: the $A_2$ minimal model, which corresponds to the potential
$$
W=\o{1}{3} ( \beta {\tilde \beta} )^3 \ \ .
\eqno(5.1)
$$
In this model, besides the identity $\Phi_0=\un$, there
is only one other chiral primary field:
$\Phi_1=(\beta{\tilde \beta})$ with $U(1)$-charge
$q=1/3$. As one can see from (2.26), the component of
the 2-form operator associated to $(\beta{\tilde \beta})$
is simply $(b{\tilde b})$. Therefore, $\int d^2w ~b(w){\tilde b}(
{\bar w})$ is the only relevant deformation which can be
used to perturb the minimal model $A_2$.
The resulting Lagrangian is then
$$
{\cal L} = {\cal L}_{\rm top} - t \int d^2w ~b(w){\tilde b}(
{\bar w})
\eqno(5.2)
$$
where ${\cal L}_{\rm top}$ is the Lagrangian for a topological
$(b,c,\beta,\gamma)$-system
as given in (2.22),
and $t$ is a dimensionful coupling constant
parametrizing its perturbation. Using the rules explained
in the previous section and in particular enforcing
the anomalous $U(1)$ charge conservation, it
is not difficult to realize that at $t=0$ the only non-vanishing
topological 3-point function for this model is
$$
c_{001} = \langle \Phi_0 \ \Phi_0 \ \Phi_1(z,{\bar z})
\rangle_{{\rm top}} = 1 \ \ .
\eqno(5.3)
$$
However, things change when $t\not = 0$.
The perturbed topological 3-point functions (see (2.25))
are in fact
given by
$$
c_{\ell_1 \ell_2 \ell_3}(t) \equiv
\left\langle \Phi_{\ell_1}(z_1,{\bar z}_1) \
 \Phi_{\ell_2}(z_2,{\bar z}_2) \
 \Phi_{\ell_3}(z_3,{\bar z}_3) \
{\rm e}^{t \int d^2w~ b(w){\tilde b}(
{\bar w})}\right\rangle_{{\rm top}}
\eqno(5.4)
$$
where $\ell_1$, $\ell_2$, $\ell_3$ can be either 0 or 1.
A simple analysis reveals that the only interesting
case is the correlation $c_{111}(t)$; all other correlators
are indeed zero because of charge conservation. To compute
$c_{111}(t)$ we expand the exponential and evaluate each
term using the bosonization rules of Section 4.
In particular, once
again because of charge conservation, all terms in this
expansion vanish except for the first-order one.
Thus we obtain
$$
\eqalign{
c_{111}(t) & = t ~ \left\langle
\Phi_{1}(z_1,{\bar z}_1) \
\Phi_{1}(z_2,{\bar z}_2) \
\Phi_{1}(z_3,{\bar z}_3) \
\int d^2w~ b(w){\tilde b}(
{\bar w})\right\rangle_{{\rm top}} \cr
& = t ~ \int d^2w ~
\left\langle \beta (z_1) \, \beta (z_2)\, \beta (z_3) \, b(w)
\right\rangle_{{\rm top}}
{}~\left\langle {\tilde{\beta}}({\bar z}_1) \,{\tilde{\beta}}
({\bar z}_2) \,{\tilde{\beta}} ({\bar z}_3) \,
{\tilde{b}}({\bar w})\right\rangle_{{\rm top}} \ \ .}
\eqno(5.5)
$$

Let us now turn to the calculation of the conformal blocks
appearing in the integrand of (5.5). We will focus just on the
holomorphic piece, since the anti-holomorphic one is simply
obtained  by complex conjugation.
First of all let us use (4.19) and (4.11)
for $n=2$ and write
$$
\eqalign{
\beta & = \exp\left[\o{1}{\sqrt{3}} \phi_0\right] ~ \varphi_1^1 \ \ , \cr
b & = \exp\left[-\o{2}{\sqrt{3}} \phi_0\right] ~ \varphi^1_{-1} \ \ .}
\eqno(5.6)
$$
Then, using (5.6) together with the definition of topological correlation
functions given at the end of Section 4, we have
$$
\eqalign{
\langle \beta(z_1) \,&  \beta(z_2) \, \beta(z_3) \, b(w) \rangle_{{\rm top}}
\cr
& \equiv
\langle 0,-3,-1| \beta(z_1) \, \beta(z_2) \, \beta(z_3) \, b(w)
|0,0,0\rangle   \cr
& = \langle 0| {\rm e}^{\left[\o{1}{\sqrt{3}} \phi_0(z_1)\right]}
{}~{\rm e}^{\left[\o{1}{\sqrt{3}} \phi_0(z_2)\right]}
{}~{\rm e}^{\left[\o{1}{\sqrt{3}} \phi_0(z_3)\right]}
{}~{\rm e}^{\left[-\o{2}{\sqrt{3}} \phi_0(w)\right]}
|0\rangle
{}~\times \cr
& ~ ~ \times
\langle-3,-1|\varphi_1^1(z_1)~\varphi_1^1(z_2)~\varphi_1^1(z_3)
{}~\varphi^1_{-1}(w)|0,0\rangle \ \ .}
\eqno(5.7)
$$
The $\phi_0$-contribution in (5.7) is immediate: the
charges exactly soak up the background anomaly $Q_0^{(2)}=-1/\sqrt{3}$ and
so we get
$$
\eqalign{
\langle 0| {\rm e}^{\left[\o{1}{\sqrt{3}} \phi_0(z_1)\right]}
& ~{\rm e}^{\left[\o{1}{\sqrt{3}} \phi_0(z_2)\right]}
{}~{\rm e}^{\left[\o{1}{\sqrt{3}} \phi_0(z_3)\right]}
{}~{\rm e}^{\left[-\o{2}{\sqrt{3}} \phi_0(w)\right]}
|0\rangle
\cr
& = ~ ~\left[(z_1-z_2)(z_1-z_3)(z_2-z_3)\right]^{\o{1}{3}}
{}~\left[(z_1-w)(z_2-w)(z_3-w)\right]^{-\o{2}{3}}
\ \ .}
\eqno(5.8)
$$
The parafermion contribution
$$
\langle-3,-1|\varphi_1^1(z_1)~\varphi_1^1(z_2)~\varphi_1^1(z_3)
{}~\varphi^1_{-1}(w)|0,0\rangle
\eqno(5.9)
$$
is also easily computed. The most efficient way is perhaps the
following: if we take into account
the identification (4.12), we see that
the fields $\varphi_1^1$ and $\varphi_{-1}^1$ are
both proportional to $\varphi^0_0 = \un$.
Moreover, the vacuum $\langle-3,-1|$
is proportional to $\langle
-2,0|$  since their dual conjugates
$|1,1\rangle$ and  $|0,0\rangle$ are equivalent because of (4.12).
Thus, (5.9) is simply the vacuum expectation value of the identity
and so it is a constant.
One can also verify this result by explicitly computing
(5.9) using for example the method of the screening charges
[38,45]. Putting everything together, we obtain
$$
\eqalign{
\langle \beta(z_1) & \, \beta(z_2) \, \beta(z_3) \, b(w) \rangle_{{\rm top}}\cr
& ~ ~\sim
\left[(z_1-z_2)(z_1-z_3)(z_2-z_3)\right]^{\o{1}{3}}
{}~\left[(z_1-w)(z_2-w)(z_3-w)\right]^{-\o{2}{3}}
\ \ .}
\eqno(5.10)
$$
Notice that this topological correlation function {\it does} depend
on the coordinates $z_i$ where the chiral fields $\beta$ are inserted.
This is not at all a surprise because (5.10) is not
a correlator of only physical fields.

The full topological correlation function $c_{111}(t)$,
which of course should be independent
of $z_i$, can now be easily computed. If we substitute (5.10)
and its complex conjugate into (5.5), we get
$$
c_{111}(t) \sim t ~ \left(|z_1-z_2|\,|z_1-z_3|\,
|z_2-z_3|\right)^{\o{2}{3}} ~\int d^2w ~
\left(|z_1-w|\,|z_2-w|\,|z_3-w|\right)^{-\o{4}{3}} \ \ .
\eqno(5.11)
$$
The integral $I$ in (5.11) is evaluated using elementary techniques
and the result is
$$
I \sim \left(|z_1-z_2|\,|z_1-z_3|\,
|z_2-z_3|\right)^{-\o{2}{3}} \ \ .
\eqno(5.12)
$$
Absorbing all numerical factors of (5.10) and (5.11) into a
rescaling of $t$,
we can conclude that
$$
c_{111}(t) = t
\eqno(5.13)
$$
Notice that only after doing the integral over $d^2w$, the correlation
function (5.11) becomes independent of the coordinates $z_i$ of
the physical fields, as it should.
In the minimal model $A_2$, $c_{111}(t)=t$ and $c_{001}(t)=1$
are the only non vanishing topological 3-point functions.

As it is well known, the 2-point function $\eta_{\ell_1\ell_2}(t)
\equiv c_{0\ell_1\ell_2}(t)$ serves as a metric in the space
of coupling constants. It can be proven on general grounds
[23] that this metric is flat and hence there exist
special flat coordinates in which it is
constant. For the minimal model
$A_2$ we simply have
$$
\eta_{00}(t) = \eta_{11}(t) = 0 ~ ~ , ~ ~ \eta_{01}(t)=1\ \ .
\eqno(5.14)
$$
Since $\eta(t)$ is not only flat but also constant,
the parameter
$t$ in (5.2) is a flat coordinate.

This example is however too trivial to let us reach any
conclusion, and one should test our methods in more
complicated cases. This can be easily done and in several
non-trivial examples for single minimal models we have
checked that the parameters entering the deformed LG potential
in our formulation are indeed flat coordinates. This is to be
contrasted with the usual formulation where the parameters
of the deformed LG potential are {\it not} flat coordinates
but are related to these by more or less complicated transformations.
The origin of this difference is that we compute the perturbed correlation
functions without using the residue pairing defined
by the perturbed potential [23] and
stay in the context of perturbed conformal field theory
Thus we can always maintain in a natural way a frame of
flat coordinates.

We have verified these properties also in the case of
the torus, which can be described by the tensor
product of three minimal models $A_2$ deformed
by its marginal operator. Therefore, the potential we should consider
is
$$
W = \o{1}{3}(\beta_x \tilde \beta_x)^3 +\o{1}{3}(\beta_y \tilde \beta_y)^3 +
\o{1}{3}(\beta_z \tilde \beta_z)^3 -t
\,(\beta_x \tilde \beta_x)(\beta_y \tilde \beta_y)(\beta_z \tilde \beta_z)
\eqno(5.15)
$$
The dimensionless parameter $t$ in (5.15)
is (related to) the modulus of the torus.
The chiral ring of the tensor product of three
$A_2$ minimal models  is generated by
$\Phi_0=\un$,
$(\Phi_1, \Phi_2, \Phi_3)=(\beta_x \tilde \beta_x, \beta_y \tilde \beta_y,
\beta_z \tilde \beta_z)$, $\Phi_4=\beta_x \tilde \beta_x\beta_y \tilde
\beta_y$, $\Phi_5=\beta_y \tilde \beta_y\beta_z \tilde \beta_z$,
$\Phi_6=\beta_x \tilde \beta_x\beta_z \tilde \beta_z$,  and $\Phi_7=\beta_x
\tilde \beta_x\beta_y \tilde \beta_y\beta_z \tilde \beta_z$. $\Phi_7$ is a
marginal operator
and is the one  appearing in (5.15) as a deformation.
The Lagrangian corresponding to
(5.15) is
$$
{\cal L} = {\cal L}_{\rm top} - t \, \int \Phi_7^{(2)}
\eqno(5.16)
$$
where ${\cal L}_{\rm top}$ is the  Lagrangian for three
topological $(b,c,\beta,\gamma)$-systems as in (2.22), and
$\Phi_7^{(2)}$
is the 2-form associated to $\Phi_7$. According to (2.29), we have
$$
\int \Phi_7^{(2)}=\int d^2 w \left[ (b_x \beta_y \beta_z +
b_y \beta_x \beta_z + b_z \beta_x \beta_y) (w) \times (\hbox{c.c})
\right] \ \ .\eqno(5.17)
$$

The simplest way to check whether the parameter $t$ in (5.16)
is a flat coordinate or not, is to compute the topological
metric
$$
\eta_{ij}(t) = \left\langle \Phi_i \,\Phi_j \,{\rm e}^{t\int \Phi_7^{(2)}
} \right\rangle_{\rm top}
\eqno(5.18)
$$
and see whether it is constant or not. To this end it is enough
to consider one component, for example
$$
\eta_{07}(t) = \left\langle \Phi_0 \,\Phi_7 \,{\rm e}^{t\int \Phi_7^{(2)}
} \right\rangle_{\rm top} \ \ .
\eqno(5.19)
$$
By looking at the $U(1)$-charges of the operators in (5.19),
it is easy to realize that when expanding the exponential,
only terms with $3n$ insertions of $\int \Phi_7^{(2)}$
will satisfy charge conservation. Thus, (5.19) can be
rewritten as
$$
\eta_{07}(t) = \sum_{n=0}^{\infty}\o{1}{(3n)!} a_{3n}\,t^{3n}
\eqno(5.20)
$$
where
$$
a_{3n} = \left\langle \Phi_0 \,\Phi_7 \left(\int \Phi_7^{(2)}
\right)^{3n}\right\rangle_{\rm top}  \ \ .
\eqno(5.21)
$$
It is immediate to see that $a_0=1$. The first non-trivial contribution is
$a_3$ which, spelled out in detail, is
$$
a_3 = \int d^2w_1 \int d^2w_2 \int d^2w_3
\,f(u,w_1,w_2,w_3){\bar f}({\bar u},{\bar w}_1,{\bar w}_2,{\bar w}_3)
\eqno(5.22)
$$
where
$$
f(u,w_1,w_2,w_3) = \left\langle (\beta_x\beta_y\beta_z)(u)
\prod_{i=1}^3 \left[(b_x \beta_y \beta_z +
b_y \beta_x \beta_z + b_z \beta_x \beta_y) (w_i)\right]\right\rangle_{\rm top}
\eqno(5.23)
$$
and ${\bar f}$ is its complex conjugate. To compute $f$, we split
the r.h.s. of (5.23) into a sum of factorized correlation
functions for each of the three minimal models and enforce on them
(anomalous) conservation of the $U(1)$ charges.
During this process the $b$ fields in (5.23) must be suitably rearranged
and proper minus signs arise from their anticommutation relations.
After some straightforward algebra, it turns out that
$f=0$, which implies
$$
a_3=0
\eqno(5.24)
$$
Actually it is very easy to generalize this result to all higher-order
coefficients, and eventually conclude that
$$
\eta_{07}(t) = 1
\eqno(5.25)
$$
The other entries of the metric $\eta_{ij}(t)$ can be computed
in a similar way and all of them turn out to be constants.
Thus the parameter $t$ in (5.16) is a flat coordinate.

We want to emphasize that instead, in the standard LG formulation
of the torus described by the potential
$$
W = \o{1}{3} X^3 + \o{1}{3} Y^3 + \o{1}{3} Z^3 - s XYZ
\eqno(5.26)
$$
the topological correlation $\langle \Phi_0 \,\Phi_7\rangle(s)$ {\it is}
a non-trivial function of the LG parameter $s$, which therefore
cannot be a flat coordinate
\footnote{$^{7}$}{It turns out that $\langle \Phi_0 \,\Phi_7\rangle(s) =
\left[F(\o{1}{3},\o{1}{3};\o{2}{3};s^3)\right]^{-2}$, where $F$ is the
hypergeometric function.}.

It is interesting now to investigate the
relation between  $s$ and $t$. On general grounds [25], it is possible
to show that
$$
s(t) = \o{c_{111}(t)}{c_{123}(t)}
\eqno(5.27)
$$
where
$$
c_{ijk}(t) = \left\langle \Phi_i \,\Phi_j\, \Phi_k \,{\rm e}^{t\int
\Phi_7^{(2)}
} \right\rangle_{\rm top} \ \ .
\eqno(5.28)
$$
In our formulation it is easy to compute
these perturbed 3-point functions as a power
series in the flat coordinate $t$.
Let us briefly see how $c_{123}(t)$ is evaluated.
Expanding the exponential and looking for terms which satisfy charge
conservation, we get
$$
c_{123}(t) = \sum_{n=0}^{\infty}\o{1}{(3n)!} {\tilde a}_{3n}\,t^{3n}
\eqno(5.29)
$$
where
$$
{\tilde a}_{3n} = \left\langle \Phi_1\,\Phi_2 \,\Phi_3 \left(\int \Phi_7^{(2)}
\right)^{3n}\right\rangle_{\rm top}  \ \ .
\eqno(5.30)
$$
It is immediate to see that ${\tilde a}_0=1$. The next contribution is
$$
{\tilde a}_3 = \int d^2w_1 \int d^2w_2 \int d^2w_3
\,g(x,y,z,w_1,w_2,w_3){\bar g}({\bar x},{\bar y},{\bar z},
{\bar w}_1,{\bar w}_2,{\bar w}_3)
\eqno(5.31)
$$
where
$$
g(x,y,z,w_1,w_2,w_3) = \left\langle \beta_x(x)\beta_y(y)\beta_z(z)
\prod_{i=1}^3 \left[(b_x \beta_y \beta_z +
b_y \beta_x \beta_z + b_z \beta_x \beta_y) (w_i)\right]\right\rangle_{\rm top}
\ \ .
\eqno(5.32)
$$
By splitting the r.h.s. of (5.32) into a sum of factorized
terms and using the explicit results derived earlier for each
of the three minimal models, one can prove that the integrand of (5.31)
is
$$
|x-y|^2 |x-z|^2 |y-z|^2
\prod_{i=1}^3 \left(|x-w_i|\,|y-w_i|\,|z-w_i|\right)^{-\o{4}{3}}
\eqno(5.33)
$$
Using (5.12) and rescaling $t$ to absorb all numerical constants, we
conclude that
${\tilde a}_3=1$, and hence
$$
c_{123}(t) = 1 +\o{1}{6}t^3 + O(t^6)
\eqno (5.34)
$$
Similarly one can check that
$$
c_{111}(t) = t + O(t^7)
\eqno(5.35)
$$
so that from (5.27) it follows
$$
s(t) = t - \o{1}{6}t^4 + O(t^7)
\eqno(5.36)
$$
These are precisely the first terms in the power series
expansion of the solution of the schwarzian differential equation
$$
\{s;t\} = - \o{1}{2} \o{(8+s^3)}{(1-s^3)^2} (s')^2 s
\eqno(5.37)
$$
where
$$
\{s;t\} = \o{s'''}{s'} - \o{3}{2} \left(\o{s''}{s'}\right)^2
$$
and the primes denote $t$-derivatives.
It has been recently shown [25,26,29] that (5.37) is equivalent to
the requirement that $t$ be a flat coordinate. Our methods
provide automatically the solution to (5.37) as a power series.
As pointed out at the beginning of this section there is no
obstruction to extend our techniques to the calculation of topological
correlators in $c >3$ models. The actual calculation of these
correlators is postponed to future work.
\vfill\eject
\centerline{\bf Appendix A}
\vskip 0.4truecm
\centerline{\bf Landau-Ginzburg action and transformation rules in
component formalism}
\vskip 0.4truecm
In this appendix we present the explicit form of the Landau-Ginzburg
lagrangian in component formalism and use the rheonomic approach
to find the N=2 supersymmetry transformations.
In the notations of [21],
we write the following curvatures
$$
\eqalign{
T^a &={\cal D}V^a -\o{\rm i}{2} \bar \xi \wedge \gamma^a \xi \ \ ,\cr
\rho &={\cal D}\xi \ \ ,\cr
F&=dA -i\bar \xi \wedge \xi \ \ ,\cr
R^{ab}&=d\omega^{ab}\ \ ,\cr}\eqno({\rm A.1})
$$
where $V^a$ is the zweibein,
$\xi$ is the gravitino one-form, $A$ is the $U(1)$ connection,
$\omega^{ab}$ is the spin connection and $\cal D$ is the Lorentz
covariant derivative. The gravitino $\xi$ is a
Dirac spinor. In general we can write
$$
\xi=e^{-i \pi /4} \twovec {\zeta}{\tilde \zeta}\eqno({\rm A.2})
$$
with $\zeta \ne \zeta^*$, $\tilde \zeta^* \ne \zeta$.
More precisely, if we set
$$\eqalign{
e^{\pm}&=\o{1}{2}(V^0 \pm V^1)\ \ , \cr
\omega^{ab}&=\epsilon^{ab}\omega \ \ ,\cr}\eqno({\rm A.3})
$$
we obtain
$$
T^a=dV^a -\omega^{ab}\wedge V^b-\o{\rm i}{2}\bar \xi \wedge \gamma^a \xi \ \ ,
\eqno({\rm A.4})$$
or
$$
T^{\pm}=de^{\pm} \pm\omega \wedge e^\pm -\o{\rm i}{2}
\bar \xi \wedge \gamma^\pm \xi \ \ ,\eqno({\rm A.5})
$$
where $\gamma^\pm\equiv\o{1}{2} (1\pm\gamma_3)$. Using (A.2), we have
$$\eqalign{
T^{+}&=de^{+} + \omega \wedge e^{+} -\o{\rm i}{2}\zeta^* \wedge \zeta \ \ ,\cr
T^{-}&=de^{-} + \omega \wedge e^{-} -\o{\rm i}{2}\tilde\zeta^* \wedge
\tilde\zeta \ \ .\cr}\eqno({\rm A.6})
$$
Similarly, we get
$$
F=dA - \zeta^* \wedge \tilde \zeta + \tilde \zeta^* \wedge \zeta \ \ .
\eqno({\rm A.7})$$
Following the general recipes of the rheonomic procedure [21], we write the
background Maurer-Cartan equations
$$
\eqalign{
de^+ &+ \omega\wedge e^+ -\o{\rm i}{2} \zeta^{+} \wedge \zeta^{-}=0\ \ ,\cr
de^- &- \omega\wedge e^- -\o{\rm i}{2} \tilde\zeta^{+} \wedge
\tilde\zeta^{-}=0\ \ ,\cr
d \zeta^+ &+\o{1}{2} \omega \wedge \zeta^+=0\ \ ,\cr
d \tilde\zeta^+ &-\o{1}{2} \omega \wedge \tilde\zeta^+=0\ \ ,\cr
d\omega&=0\ \ ,\cr
dA &- \zeta^- \wedge \tilde \zeta^+ +\tilde \zeta^+ \wedge \zeta^-=0 \ \ ,\cr}
\eqno({\rm A.8})$$
where we have set $\zeta^-=\zeta$ and $\zeta^+=\zeta^*$.
Using these notations, we can write the general form of the LG
lagrangian in components.
{}From the Bianchi identities $d^2 X^i=d^2 \psi^i=d^2 \psi^{i^*}=0$ and from
(A.8) one derives the following rheonomic parametrizations
$$
\eqalign{
d\Xi&=\part_+\Xi e^+ + \part_-\Xi e^- +\psi^i \zeta^- +\tilde \psi^i \tilde
\zeta^- \ \ ,\cr
d\Xis&=\part_+\Xis e^+ +\part_-\Xis e^-  -\psi^{i^*} \zeta^+
-\tilde \psi^{i^*} \tilde \zeta^+ \ \ ,\cr
d \psi^i &=\part_+\psi^i e^+ +\part_-\psi^i e^--\o{\rm i}{2} \part_+ \Xi
\zeta^+ +\eta^{ij^*}\part_{j^*} \bar W \zeta^- \ \ ,\cr
d\tilde \psi^i &=\part_+ \tilde \psi^i e^+ +\part_- \tilde \psi^i e^-
-\o{\rm i}{2}\part_-\Xi \tilde \zeta^+ - \eta^{ij^*} \part_{j^*} \bar W
\zeta^- \ \ ,\cr}\eqno({\rm A.9})
$$
where $\Xi, \Xis=(\Xi)^*$ are complex coordinates in a flat K\"ahler
manifold, $\eta^{ij^*}$ is the flat metric and $\psi^i$, $\tilde \psi^i$
are the complex spin-$\o{1}{2}$ fermionic partners of $\Xi$'s.
The parametrizations of $d\psi^{i^*}$ and $
d \tilde \psi^{i^*}$ are obtained by complex conjugation.
Using standard techniques, one finds that the action, from which
(A.9) follow as field equations in the vertical directions, is
$$
S=\int {\cal L}\eqno({\rm A.10})
$$
where
$$
\eqalign{
{\cal L} =&~\eta_{ij^*}(d\Xi - \psi^i \zeta^- -\tilde \psi^i \tilde
\zeta^-)\wedge(\Pi^{j^*}_+ e^+ - \Pi_-^{j^*} e^-)\cr
& +\eta_{j^*i}(d\Xjs + \psi^{j^*}\zeta^+ + \tilde \psi^{j^*} \tilde
\zeta^+)\wedge(\Pi_+^i e^+ -\Pi_-^i e^-)\cr
& +\eta_{ij^*}(\Pi_+^i\Pi_-^{j^*} + \Pi^i_-\Pi_+^{j^*})e^+\wedge
e^- -(4{\rm i} \eta_{ij^*} \psi^i d\psi^{j*} +\o{\rm i}{2} \psi^k
\partial_k W \tilde \zeta^+)\wedge e^+ \cr
 &+ (4{\rm i} \eta_{ij^*} \tilde \psi^i d\tilde\psi^{j*} +\o{\rm i}{2}
\tilde\psi^k
\partial_k W \zeta^+)\wedge e^- \cr
&+\o{\rm i}{2} \psi^{k^*}
\partial_{k^*} \bar W \tilde\zeta^-\wedge e^+
-\o{\rm i}{2}
\tilde\psi^{k^*}
\partial_{k^*} \bar W \zeta^-\wedge e^- \cr
& +(  8 \psi^i \tilde \psi^j \part_i \part_j W +8
\tilde \psi^{i^*} \tilde \psi^{j^*} \part_{i^*} \part_{j^*} \bar W
+8 \eta^{ij^*} \part_i W \part_{j^*} \bar W) e^+ \wedge e^- \cr
& + d\Xjs \wedge \psi^i \zeta^- \eta_{i j^*} - d\Xjs
\wedge \tilde \psi^i \tilde \zeta^- \eta_{i j^*}\cr
& + dZ^{i} \wedge \psi^{j^*} \zeta^+ \eta_{i j^*}
- dZ^{i} \wedge \tilde \psi^{j^*} \tilde\zeta^+ \eta_{i j^*} \ \ .\cr}
\eqno({\rm A.11})$$
The lagrangian (A.11) is written in first-order formalism and the auxiliary
fields $\pi^i_\pm, \pi^{i^*}_\pm$ can be eliminated through their
equations of motion: $\Pi_{\pm}^i=\part_\pm \Xi$ and
$\Pi_{\pm}^{i^*}=\part_\pm \Xis$.
The lagrangian (2.29) of Section 2 is obtained from (A.11) by
restricting it to the bosonic surface (namely discarding all terms
containing the gravitino forms $\zeta^\pm ,\tilde \zeta^\pm$) and
substituting back the above mentioned equations for $\Pi_\pm^{i,i^*}$.
Form the curvature parametrization (A.9), we easily recover the N=2
supersymmetry
transformations
$$\eqalign{ \delta \Xi \, &= - \varepsilon^{-} \posi -
{\tilde \varepsilon}^{-} \posib \ \ ,\cr
\delta \Xis \, &= \varepsilon^{+} \posis +
{\tilde \varepsilon}^{+} \posisb \ \ ,\cr
\delta \posi \, & = -\o {\rm i}{2} \pa \Xi \, \varepsilon^{+} +
\eta^{ij^{\star}}
\pa_{j^{\star}} {\bar W} \, {\tilde \varepsilon}^{-}\ \ ,\cr
\delta \posib \, & =-\o {\rm i}{2} \pab \Xi \,
{\tilde \varepsilon}^{+}-
\eta^{ij^{\star}}
\pa_{j^{\star}} {\bar W} \, \varepsilon^{-} \ \ ,\cr
\delta \posis \, & = \,  \o {\rm i}{2}
\pa \Xis \, \varepsilon^{-} +  \eta^{ji^{\star}}
\pa_{j} W \, {\tilde \varepsilon}^{+}\ \ ,\cr
\delta \posisb \, & = \,  \o {\rm i}{2}
\pab \Xis \, {\tilde \varepsilon}^{-} -  \eta^{ji^{\star}}
\pa_{j} W \, \varepsilon^{+} \ \ ,\cr}\eqno({\rm A.12})$$
which coincide precisely with the ones in (2.31) of Section 2.
\vskip 2truecm
\centerline{\bf Appendix B}
\vskip 0.4truecm
\centerline{\bf The lagrangian for the topological $(b,c,\beta,\gamma)$-system}
\vskip 0.4truecm
In Section 2 we gave the explicit form of the descent equations (2.26),
using (2.23) as BRST tranformations. We know
that this case corresponds to the twisted stress-energy tensor
$$
\hat T_+ = T + \o{1}{2}\part J=\part c b + \gamma \part\beta
\eqno({\rm B.1})
$$
which is associated to a $(b,c,\beta,\gamma)$-system with
$\lam_\beta=0$, $\lam_b=0$. For simplicity we consider the case of only
one collection of
pseudo-ghost fields and as usual, we understand the
expressions for the tilded operators.
The main advantage of this approach is that we can
write the chiral primary fields of the N=2 theory in terms of the
{\sl local} fields appearing in the lagrangian, and hence we can
construct the representatives of the BRST cohomology in a rather
simple way. The lagrangian, however, is not a BRST commutator: we are dealing
with a topological theory of the Schwartz-type [22].

As pointed out in Section 2, there is another possibilty and one can take as
twisted stress-energy tensor the following expression
$$\eqalign{
{\hat T}_- & =T-\o{1}{2} \part J \cr
&= (1-2\lam)\,\partial  c \,
b  \, - 2\lam \, b \partial c \,
+ \, (1-2\lam) \,\gamma \, \partial
\beta  \,  - \,  2\lam  \beta \, \partial \gamma \ \ .}
\eqno({\rm B.2})
$$
This
corresponds to a commuting $(\beta,\gamma)$-system with
weights $\lam_\beta
=2\lam$, $\lam_{\gamma}=1-2\lam$, and to an anticommuting $(b,c)$-system
with weights
$\lam_b=2\lam$ and $\lam_{c}=1 -2\lam$.
In this case the BRST tranformations are
$$
\eqalign{
s\beta&=2b ~~~, ~~~ s\tilde \beta=2\tilde b \ \ ,\cr
sc&=2\gamma ~~~, ~~~ s \tilde c=2 \tilde \gamma \ \ ,\cr
sb&=0 ~~~,~~~ s\tilde b=0 \ \ ,\cr
s\gamma &=0 ~~~,~~~ s\tilde \gamma=0 \ \ .\cr}\eqno({\rm B.3})
$$
Eq.s (B.3) are obtained from the supersymmetry tranformations (2.8) by
setting $\epsilon^+=\tilde \epsilon^+=0$ and choosing as BRST
parameter $\theta=\sqrt{2} \epsilon^-=\sqrt{2} \bar\epsilon^-$
This corresponds to taking
$$
Q=\o{1}{\sqrt{2}} \left(\oint G^-(z)dz +\oint \bar G^-({\bar z})d \bar z
\right )\eqno({\rm B.4})
$$
as BRST charge.
The stress-energy tensor (B.2) is a BRST commutator, namely
$$
\hat T=s\left [-(\lam -\o{1}{2}) c \part \beta -\lam \beta \part c
\right ]\eqno({\rm B.5})
$$
where, as usual, we have defined $s\phi=[Q,\phi]$ for a generic field
$\phi$.
Now we want to show
that the twist (B.2) can be seen directly at the
lagrangian level, or equivalently that we can write
the lagrangian (
including the interaction term) as a BRST commutator. If this is the case, then
we have a topological theory of Witten-type [22].
To understand this point, we first recall that the supercurrent $G^+$,
as shown in (2.9), is
actually composed of a $G^+_z$ term and a $G^+_{\bar z}$
term. The latter is zero on shell, but it
has to be taken into account in defining the superconformal
transformations [37]. Keeping this in mind,
we recognize that
$$\eqalign{
{\cal L} &= s (- G^+_{\bar z}- \bar G^+_z) \cr
& =s [-(\lam -\um) c\bar\part \beta
-\lam \beta \bar \part c + \um V \tilde b
\tilde \part \tilde V
-(\lam -\um) \tilde c \part \tilde\beta
-\lam \tilde \beta \part \tilde c - \um \tilde V  b \part V ]
\ \ .\cr}  \eqno({\rm B.6})
$$
Using (B.5), we get
$$
{\cal L}=\left[-2\lam \beta \bar \part \gamma + (1-2\lam) \gamma \bar
\part \beta + (1-2 \lam)\bar \part b c -2\lam b \bar \part c +
b \tilde b \part V \bar \part \tilde V\right] +\hbox{c.c} \eqno({\rm B.7})
$$
which is the expected lagrangian for the twisted $\lambda$'s.
It is suggestive to point out that, if we define
$$\eqalign{
G^+ &=-(\lam -\o{1}{2})c ~d\beta
-\lam\beta ~dc +\o{1}{2} V \tilde b \tilde \part
\tilde V d\bar z \ \ ,\cr
\bar G^+ &=-(\lam -\o{1}{2})\tilde c ~d\tilde\beta
-\lam\tilde\beta ~d\tilde c -\o{1}{2} \tilde V b  \part
V dz \ \ ,\cr}
\eqno({\rm B.8})
$$
where for a generic pseudo-ghost field $\phi$
$$
d\phi=\part \phi dz +\bar \part \phi d \bar z \eqno({\rm B.9})
$$
denotes the space-time part of the rheonomic parametrizations
(that is we disregard the gravitino contributions), we get
$$
S=\int {\cal L}_{\rm top} ~dz \wedge d \bar z=\int dz \wedge s ( G^+)
+ \int  s ( \bar G^+)\wedge d\bar z\ \ .\eqno(
{\rm B.10})
$$
Finally, if we regard (B.7) as
a topological lagrangian
without special requirements on the interaction term, we can ask
when this model
defines a conformal field theory. The answer to this question is almost
obvious, even if the calculation is a little different: only for
quasi-homogeneous potential $V(\beta)$ with $2\lambda=\omega$
we get a conformal field theory.
\vfill\eject
\centerline{\bf References}
\vskip 0.2cm

\def\title#1{``{#1}''}
\def\j#1#2#3#4{ #1  {\bf {#2}} $(#3)~#4$}
\def\np{Nucl. Phys.}
\def\pl{Phys. Lett.}

\item{[1]}{For a review see M.B. Green, J. Schwarz and E. Witten,
\title{Superstring theory}, Cambridge University Press, 1987}.
\item{[2]}{M. Ademollo, L. Brink, A. D'Adda, R. D'Auria, E. Napolitano,
S. Sciuto, E. del Giudice, P. di Vecchia, S. Ferrara,
F. Gliozzi, R. Musto and R. Pettorino, \j{\pl}{62B}{1976}{105}}.
\item{[3]}{A. Sen, \j{\np}{B278}{1986}{289} and \j{\np}{B284}
{1987}{423}};
\item{}{T. Banks, L.J. Dixon, D. Friedan and E. Martinec, \j{
\np}{B299}{1988}{613} and \j{\np}{B307}{88}{93}}.
\item{[4]}{D. Gepner, \j{\np}{B296}{1988}{757}
and \j{\pl}{199B}{1987}{380}}.
\item{[5]} W. Lerche, C. Vafa and N.P. Warner, \j{\np}{B324}{1989}{427};
\item{} B. Greene, C. Vafa and N.P. Warner, \j{\np}{B324}{1989}{371}.
\item{[6]} T. Eguchi, H. Ouguri, A. Taormina and S. K. Yang, \j{\np}
{B315}{1989}{192}.
\item{[7]}{P. Candelas, C.T. Horowitz, A. Strominger and E. Witten,
\j{\np}{B298}{1988}{493}}.
\item{[8]} P. Candelas, A.M. Dale, C.A. Lutken and R. Schimmrick,
\j{\np}{B298}{1988}{493}.
\item{[9]} M. Linker and R. Schimmrick,  \j{\pl}{208B}{1988}{216} and
\j{\pl}{215B}{1988}{681}.
\item{[10]}C.A. Lutken and G.C. Ross, \j{\pl}{213B}{1988}{152}.
\item{[11]} P. Zoglin, \j{\pl}{218B}{1989}{444}.
\item{[12]}B. de Wit and A. Van Proeyen, \j{\np}{B245}{1984}{89}.
\item{[13]}E. Cremmer, C. Kounnas, A. Van Proeyen, J.P. Derendinger,
S. Ferrara, B. de Wit and L. Girardello, \j{\np}{B250}{1985}{385}.
\item{[14]} B. de Wit, P.G. Lauwers and A. Van Proeyen, \j{
\np}{B255}{1985}{560}.
\item{[15]}L. Castellani, R. D'Auria and S. Ferrara, \j{
\pl}{241B}{1990}{57}; \j{Class. and Quantum Grav.}{1}{1990}{163};
\item{}R. D'Auria, S. Ferrara and P. Fr\`e, \j{\np}
{B359}{1991}{705}.
\item{[16]} A. Strominger, \j{Comm. Math. Phys.}{133}{1990}{163}.
\item{[17]} S. Cecotti, S. Ferrara and L. Girardello, \j{Int. Jour. Mod
. Phys}{A10}{1989}{2475} and \j{\pl}{213B}{1988}{443}.
\item{[18]} S. Ferrara and A. Strominger,  Strings 89, Eds.
R. Arnowitt, R. Bryan, M. Duff, D. Nanopulos and C. Pope, World
Scientific, Singapore, 1989.
\item{[19]}{P. Candelas and X. de la Ossa, \j{\np}{B342}{1990}{246}.
\item{[20]} L. J. Dixon, V. Kaplunowski and J. Louis, \j{
\np}{B329}{1990}{27}.
\item{[21]} For a review see L. Castellani, R. D' Auria and P. Fr\`e,
\title{Supergravity and
superstrings: a geometric prospective}, World Scientific, Singapore
(1991).
\item{[22]}For a review see D. Birmingham, M. Blau and M. Rakowski, \j{
Phys. Rep.}{209}{1991}{129}.
\item{[23]} R. Dijkgraaf, E. Verlinde and H. Verlinde, \j
{\np}{B352}{1991}59.
\item{[24]} B. Block and A. Varchenko, Prepr. IASSNS-HEP-91/5.
\item{[25]} E. Verlinde and N.P. Warner, \j{\pl}{269B}{1991}{96}.
\item{[26]} A. Klemm, S. Theisen and M. Schmidt, Prepr. TUM-TP-129/91,
KA-THEP-91-00, HD-THEP-91-32.
\item{[27]} S. Ferrara, J. Louis, Prepr. CERN-TH-6334/91.
\item{[28]} A. Ceresole, R. D'Auria, S. Ferrara, W. Lerche and J.
Louis, CERN-TH-6441/92.
\item{[29]} Z. Maassarani, Prepr. USC-91/023.
\item{[30]} E. Witten, \j{Comm. Math. Phys.}{118}{1988}{411} and \j{\np}
{B340}{1990}{281};
\item{} T. Eguchi and S.K. Yang \j{Mod Phys Lett}{A5}{1990}{1693}
\item{[31]} L. Baulieu and E. M. Singer, \j{Comm. Math. Phys.
}{125}{1989}{125}.
\item{[32]} E. Witten, Prepr. IASSNS-HEP-91/83.
\item{[33]} For a review see A. Pasquinucci, Ph.D. Thesis SISSA/EP
1990.
\item{[34]} S. Cecotti, L. Girardello and A. Pasquinucci, \j{\np}
{B328}{1989}{701} and \j{Int. J. Mod.
Phys.} {A6}{1991}{2427}.
\item{[35]} N.P. Warner, Lectures at Trieste Spring school 1988, World
Scientific, Singapore;
\item{} E. Martinec, \j{\pl}{217B}{1989}{431};
\item{} For a review see also \title{Criticality, Catastrophe and
Compactification}, V.G. Knizhnik memorial volume, 1989 .
\item{[36]} V.I. Arnold, S.M. Gusein-Zade, A.N. Varchenko,
\title{Singularities of differentiable maps}, Vol I, II Birk\"auser,
Boston.
\item{[37]} P. Fr\`e, F. Gliozzi, R. Monteiro and A. Piras,
\j{Class. and Quantum Grav}{8}{1991}{1455.}
\item{[38]} M. Frau, J.G. McCarthy, A. Lerda, S. Sciuto and J.
Sidenius, \j{\pl}{254B}{1991}{381} and \j{\pl}{245B}{1990}{453}.
\item{[39]} G. Mussardo, G. Sotkov and M. Stanishkov, \j{Int. J.
Mod. Phys.}{A4}{1986}{1135}.
\item{[40]} P. Griffiths and J. Harris, \title{Principles of algebraic
geometry} ed.s John Wiley and sons.
\item{[41]} C. Vafa, \j{Mod. Phys. Lett.}{A6}{1991}{337}.
\item{[42]} M.T. Grisaru, A. Lerda, S. Penati and D. Zanon, \j{\np}
{B342}{1990}{564} and \j{\pl}{234B}{1990}{88}.
\item{[43]} A. B. Zamolodchikov, \j{JEPT Lett.}{43}{1986}{730} and
\j{Sov. J. Nucl. Phys.}{46}{1987}{1090}.
\item{[44]} E. Gava and M. Stanishkov, \j{Mod. Phys. Lett.
}{27}{1990}{2261}.
\item{[45]} G. Felder, \j{\np}{B317}{1989}{215} and \j{\np}{B324}
{1989}{548E};
\item{}V.S. Dotsenko and V.A. Fateev, \j{\np}{B240}{1984}{312}.
\vfill\eject

\nopagenumbers
\hskip 9cm \vbox{\hbox{SISSA 28/92/EP}
\hbox{ITP--SB--92--7}
\hbox{IFUM 416/FT}
\hbox{March 1992}}
\vskip 0.4cm
\centerline{\bf TOPOLOGICAL FIRST-ORDER SYSTEMS}
\centerline{\bf WITH}
\centerline{{\bf LANDAU-GINZBURG INTERACTIONS}\footnote{$^*$}{
Work supported in part by Ministero dell'Universit\`a e della Ricerca
Scientifica e Tecnologica, and by NSF grants PHY 90-08936}}
\vskip 0.3cm
\centerline{\bf Pietro Fr\`e }
\vskip 0.1cm
\centerline{\sl SISSA - International School for Advanced Studies}
\centerline{\sl Via Beirut 2, I-34100 Trieste, Italy}
\vskip 0.1cm
\centerline{\sl and I.N.F.N. sezione di Trieste}
\vskip 0.2cm
\centerline{\bf Luciano Girardello }
\vskip 0.1cm
\centerline{\sl Dipartimento di Fisica, Universit\'a di Milano}
\centerline{\sl Via Celoria 16, I-20133 Milano, Italy}
\vskip 0.1cm
\centerline{\sl and I.N.F.N. sezione di Milano }
\vskip 0.2cm
\centerline{{\bf Alberto Lerda}\footnote{$^1$}{
Also at
{\sl Dipartimento di Fisica Teorica, Universit\'a di Torino,
Via P. Giuria 1, I-10125 Torino, Italy and I.N.F.N. sezione di Torino.}
}}
\vskip 0.1cm
\centerline{\sl Institute for Theoretical Physics, S.U.N.Y. at Stony Brook}
\centerline{\sl Stony Brook, N.Y. 11794, U.S.A.}
\vskip 0.2cm
\centerline{\bf Paolo Soriani}
\vskip 0.1cm
\centerline{\sl SISSA - International School for Advanced Studies}
\centerline{\sl Via Beirut 2, I-34100 Trieste, Italy}
\vskip 0.1cm
\centerline{\sl and I.N.F.N. sezione di Trieste}
\vskip 0.3cm
\centerline{{\bf Abstract}}
\vskip 0.2cm
We consider the realization of N=2 superconformal models
in terms of free first-order $(b,c,\beta,\gamma)$-systems, and show that
an arbitrary Landau-Ginzburg interaction with quasi-homogeneous
potential can be introduced without spoiling the (2,2)-superconformal
invariance. We discuss the topological twisting and the
renormalization group properties of these theories,
and compare them to the conventional
topological Landau-Ginzburg models. We show that in our
formulation the parameters multiplying deformation terms in the
potential are flat coordinates. After properly bosonizing
the first-order systems, we are able to make explicit calculations of
topological correlation functions as power series in these flat coordinates
by using
standard  Coulomb gas techniques. We retrieve known results for the
minimal models and for the torus.

\bye